\documentclass[12pt]{article}

\usepackage{fixltx2e}
\usepackage{textcomp}
\usepackage{fullpage}
\usepackage{amsfonts}
\usepackage{verbatim}
\usepackage[english]{babel}
\usepackage{pifont}
\usepackage{color}
\usepackage{setspace}
\usepackage{lscape}
\usepackage{indentfirst}
\usepackage[normalem]{ulem}
\usepackage{booktabs}
\usepackage{natbib}
\usepackage{float}
\usepackage{latexsym}
\usepackage{url}
\usepackage{hyperref}
\usepackage{epsfig}
\usepackage{graphicx}
\usepackage{amssymb}
\usepackage{amsmath}
\usepackage{bm}
\usepackage{array}
\usepackage{mhchem}
\usepackage{ifthen}
\usepackage{caption}
\usepackage{hyperref}
\usepackage{amsthm}
\usepackage{amstext}

\usepackage{graphicx}
\usepackage{multirow}
\usepackage{breqn}
\renewcommand{\tableofcontents}{}

\bibpunct{(}{)}{;}{a}{}{,}  % this is a citation format command for natbib

\begin{document}
%\begin{flushright}
%Version dated: \today
%\end{flushright}
%\bigskip
%\noindent RH:  Relaxed Drift Diffusion
% put in your own RH (running head)
% for POVs the RH is always POINT OF VIEW

\bigskip
\medskip
\begin{center}

% Insert your title:
\noindent{\Large \bf Understanding Past Population Dynamics: Bayesian Coalescent-Based Modeling with Covariates}
\bigskip

% We don't use a special title page; the author information is entered 
% like any other text.

% FOOTNOTES: We don't allow them in the manuscript, except in
% tables. Don't include any footnotes in the text.

\noindent {\normalsize \sc 
Mandev S. Gill$^1$, Philippe Lemey$^{2}$, Shannon N. Bennett$^{3}$, Roman Biek$^{4}$ \\ 
and Marc A. Suchard$^{1,5,6}$}\\[2ex]
\noindent {\small \it 
$^1$Department of Biostatistics, Jonathan and Karin Fielding School of Public Health, University
  of California, Los Angeles, United States\\
  $^2$Department of Microbiology and Immunology, Rega Institute, KU Leuven, Leuven, Belgium \\
  $^3$Department of Microbiology, California Academy of Sciences, San Francisco, United States \\
  $^4$Institute of Biodiversity, Animal Health and Comparative Medicine, University of Glasgow,
  Glasgow, United Kingdom
  $^5$Department of Biomathematics, David Geffen School of Medicine at UCLA, University of California,
  Los Angeles, United States \\
  $^6$Department of Human Genetics, David Geffen School of Medicine at UCLA, Universtiy of California,
  Los Angeles, United States}\\
\end{center}
\medskip

%\vspace{1in}
%\clearpage
\begin{abstract}
Effective population size characterizes the genetic variability in a population and is a
parameter of paramount importance in population genetics.  Kingman's coalescent process
enables inference of past population dynamics directly from molecular sequence data, and
researchers have developed a number of flexible coalescent-based models for Bayesian nonparametric
estimation of the effective population size as a function of time.  Major goals of demographic
reconstruction include identifying driving factors of
effective population size, and understanding the association between the effective population size
and such factors.  Building upon Bayesian nonparametric
coalescent-based approaches, we introduce a flexible framework that incorporates time-varying
covariates that exploit Gaussian Markov random fields to achieve temporal smoothing of effective population size trajectories.
To approximate the posterior distribution, we adapt efficient Markov chain Monte Carlo
algorithms designed for highly structured Gaussian models.
Incorporating covariates into the demographic inference framework enables the modeling
of associations between the effective population size and covariates while accounting for uncertainty
in population histories.  Furthermore, it can lead to more precise estimates of population dynamics.
We apply our model to four examples.  We reconstruct the demographic history
of raccoon rabies in North America and find a significant association with the spatiotemporal
spread of the outbreak.  Next, we examine the effective population size trajectory of
the DENV-4 virus in Puerto Rico along with viral isolate count data and find similar cyclic patterns.
We compare the population history of the HIV-1 CRF02_AG clade in Cameroon with HIV incidence and
prevalence data and find that the effective population size is more reflective of incidence rate.
Finally, we explore the hypothesis that the population dynamics of musk ox during the
Late Quaternary period were related to climate change.
\end{abstract}

% Points of View do not have abstracts but they should include
% Keywords.

\vspace{1.5in}

\section{Introduction}

The effective population size is an abstract
parameter of fundamental importance in population genetics, evolutionary biology and
infectious disease epidemiology.
\citet{Wright1931} introduces the
concept of effective population size
as the size of an idealized Fisher-Wright population that gains and loses genetic diversity
at the same rate as the real population under study.  The Fisher-Wright model is a classic
forward-time model of reproduction that assumes random mating, no selection or migration, and
non-overlapping generations.  Coalescent theory \citep{Kingman1982b, Kingman1982} provides a probabilistic
model for generating genealogies relating samples of individuals arising from a Fisher-Wright
model of reproduction.  Importantly, the coalescent elucidates the relationship between population genetic
parameters and ancestry.  In particular, the dynamics of the effective population size greatly inform
the shapes of coalescent-generated genealogies.  This opens the door for the inverse problem of coalescent-based
inference of effective population size trajectories from gene genealogies.
\par
While the coalescent was originally developed for constant-size populations,
extensions that accommodate a variable population size \citep{Slatkin1991, Griffiths1994, Donnelly1995}
provide a basis for estimation of the effective population size as a function of time (also called
the demographic function).  Early approaches assumed simple parametric forms for the demographic function, such
as exponential or logistic growth, and provided maximum likelihood \citep{Kuhner1998} or Bayesian
\citep{Drummond2002} frameworks for estimating the parameters that characterized the parametric forms.
However, \textit{a} \textit{priori} parametric assumptions can be quite restrictive, and finding an
appropriate parametric form for a given demographic history can be time consuming and computationally
expensive.  To remedy this, there has been considerable development of nonparametric methods
to infer past population dynamics.
\par
Nonparametric coalescent-based models typically approximate the effective population size as a piecewise
constant or linear function.  The methodology has evolved from fast but noisy models based on method of moments
estimators \citep{Pybus2000, Strimmer2001}, to a number of flexible Bayesian approaches, including
multiple change-point models \citep{Opgen-Rhein2005, Drummond2005, Heled2008}, and models that employ
Gaussian process-based priors on the population trajectory \citep{Minin2008, Gill2013, Palacios2013}.
Extending the basic methodological framework to incorporate a number of
key features, including accounting for phylogenetic error \citep{Drummond2005, Minin2008, Heled2008, Gill2013},
the ability to analyze heterochronous data \citep{Pybus2000, Drummond2005, Minin2008, Heled2008, Gill2013, Palacios2013},
and simultaneous analysis of multilocus data \citep{Heled2008, Gill2013} has hastened progress.
%(Expand into more detail, see Ho and Shapiro)
\par
In spite of all of these advances, there remains a need for further development of population dynamics inference
methodology.  One promising avenue is introduction of covariates into the inference framework.
A central goal in demographic reconstruction is to gain insights into the association between past
population dynamics and external factors \citep{Ho2011}.
For example, \citet{Lorenzen2011} combine demographic reconstructions from ancient DNA with species distribution
models and the human fossil record to elucidate how climate and humans impacted the population dynamics
of woolly rhinoceros, woolly mammoth, wild horse, reindeer, bison and musk ox during
the Late Quaternary period.  \citet{Lorenzen2011} show that changes in megafauna abundance are idiosyncratic, with
different species (and continental populations within species) responding differently to the effects
of climate change, human encroachment and habitat redistribution.
\citet{Lorenzen2011} identify climate change as the primary explanation behind the extinction of
Eurasian musk ox and woolly rhinoceros, point to a combination of climatic and anthropogenic factors
as the causes of wild horse and steppe bison decline,
and observe that reindeer remain largely unaffected by any such factors.
%For example, \citet{Campos2010} employ the
%Skyride \citep{Minin2008} and Bayesian Skyline \citep{Drummond2005} models to reconstruct the population dynamics
%of musk ox dating back to the late Pleistocene era from ancient DNA sequences.  The musk ox population was
%once widely distributed in the holarctic ecozone but is now confined to Greenland and the Arctic Archipelago,
%and \citet{Campos2010} explore potential causes of musk ox population decline.  The authors find that the
%arrival of humans into relevant areas did not correspond to changes in musk ox effective population size.
%On the other hand, \citet{Campos2010} observe that time intervals during which musk ox populations increase generally
%correspond to periods of global climatic cooling, and musk ox populations decline during warmer and
%climatically unstable periods.  Thus environmental change rather than human presence appears to
%have driven musk ox population dynamics.
Similarly, \citet{Stiller2010} examine whether climatic changes were related to
the extinction of the cave bear, and \citet{Finlay2007} consider the impact of domestication on the
population expansion of bovine species.  Comparison of external factors with past population dynamics
is also a popular approach in epidemiological studies to explore hypotheses about the spread of viruses
\citep{Lemey2003, Faria2014}.
\par
In addition to the association between past population dynamics and potential driving factors,
it is of fundamental interest is to assess the association between effective population size and
census population size \citep{Crandall1999, Liu2008, Volz2009, Palstra2012}.
For instance, \citet{Bazin2006} argue that
in animals, diversity of mitochondrial DNA (mtDNA) is not reflective of population size, whereas
allozyme diversity is.  \citet{Atkinson2008} follow up by examining whether mtDNA diversity
is a reliable predictor of human population size.  The authors compare Bayesian Skyline \citep{Drummond2005} effective
population size reconstructions with historical estimates of census population sizes and find concordance between the two
quantities in terms of relative regional population sizes.
\par
Existing methods for population dynamics inference do not incorporate covariates directly into
the model, and associations between the effective population size and potentially related factors
are typically examined in post hoc fashions that ignore uncertainty in demographic reconstructions.
We propose to fill this void by including external time series as covariates in a
generalized linear model framework.  We accomplish this task by building upon the
the Bayesian nonparametric Skygrid model of \citet{Gill2013}.  The Skygrid is a particularly
well-suited starting point among nonparametric coalescent-based models.  In most other comparable models,
the trajectory change-points must correspond to internal nodes of the genealogy, creating a hurdle
for modeling associations with covariates that are measured at fixed times.  The Skygrid
bypasses such difficulties by allowing users to specify change-points, providing a more natural
framework for our extension.  Furthermore, the Skygrid's
Gaussian Markov random field (GMRF) smoothing prior is highly generalizable and affords
a straightforward extension to include covariates.
\par
We demonstrate the utility of incorporating covariates into demographic inference on four examples.
%First, we explore the relationship between climate change and ancient horse populations and
%find a positive association between effective population size and global temperature.
First, we find striking similarities between the demographic and spatial expansion of
raccoon rabies in North America.  Second, we compare and contrast the epidemiological dynamics of dengue
in Puerto Rico with patterns of viral diversity.  Third, we examine the population history of the
HIV-1 CRF02\_AG clade in Cameroon and find that the effective population size is more reflective of
HIV incidence than prevalence.  Finally, we explore the relationship between musk ox population
dynamics and climate change during the Late Quaternary period.
%Our findings suggest that ...
Our extension to the Skygrid proves to be a useful framework for
ascertaining the association between effective population size and external covariates while
accounting for demographic uncertainty.  Furthermore, we show that incorporating covariates
into the demographic inference framework can improve estimates of effective population size trajectories,
increasing precision and uncovering patterns in the population history that integrate the covariate data
in addition to the sequence data.

\section{Methods}
\subsection{Coalescent Theory}

Coalescent theory forms the basis of our inference framework, and here we review the basic set-up.
Consider a random sample of $n$ individuals
arising from a classic Fisher-Wright population model of constant size $N_e$.  The coalescent
\citep{Kingman1982b, Kingman1982} is a stochastic process that generates genealogies relating such a sample.
The process begins at the sampling time of all $n$ individuals, $t = 0$, and proceeds
backward in time as $t$ increases, successively
merging lineages until all lineages have merged and we have reached the root of
the genealogy, which corresponds to the most recent common ancestor (MRCA)
of the sampled individuals.
The merging of lineages is called a coalescent event and
there are $n-1$ coalescent events in all.  Let $t_k$ denote the time of the $(n-k)^{\text{\small th}}$ coalescent event for $k = 1,\dots,n-1$
and $t_n = 0$ denote the sampling time.  Then for $k = 2,\dots,n$, the waiting time $w_k = t_{k-1}-t_k$
is exponentially distributed with rate $\frac{k(k-1)}{2N_e}$.
%(see Rodrigo to provide details of discrete case before moving to continuous time)
\par
Researchers have extended coalescent theory to model the effects of recombination \citep{Hudson1983},
population structure \citep{Notohara1990}, and selection \citep{Krone1997}.
We do not, however, incorporate any of these extensions here.  The relevant extensions for our development
generalize the coalescent to accommodate a variable population size \citep{Griffiths1994}
and heterochronous data \citep{Rodrigo1999b}.
The latter occurs when one samples the $n$ individuals at possibility different times.
\par
Let $N_e(t)$ denote the effective population size as a function of time, where time increases into the past.
%\citet{Griffiths1994} provide a generalization of the coalescent that allows for the effective population
%size $N_e = N_e(t)$ to change over time.
Thus, $N_e(0)$ is the effective population size at the most recent sampling time, and $N_e(t')$ is
the effective population size $t'$ time units before the most recent sampling time.
We also refer to $N_e(t)$ as the ``demographic function'' or ``demographic model.''
\citet{Griffiths1994} show that the waiting time $w_k$ between coalescent events
is given by the conditional density
\begin{equation}
P(w_k|t_k) =  \frac{k(k-1)}{2N_e(w_k + t_k)} \exp \left[- \int_{t_k}^{w_k + t_k} \frac{k(k-1)}{2N_e(t)} dt  \right] .
\end{equation}
Taking the product of such densities yields the joint density of intercoalescent waiting times, and
this fact can be exploited to obtain the probability of observing a particular genealogy given a
demographic function.
%Here, we consider a piecewise constant demographic function that changes values at pre-specified times.
%add stuff about density with heterochronous data

\subsection{Skygrid Demographic Model}
\par
The Skygrid posits that $N_e(t)$ is a piecewise constant function that can change values only at
pre-specified points in time known as ``grid points.''  Let $x_1,\dots,x_M$ denote the temporal
grid points, where $x_1 \leq x_2 \leq \dots \leq x_{M-1} \leq x_M$.
The $M$ grid points divide the demographic history timeline into $M+1$ intervals so that the
demographic function is fully specified by a vector $\boldsymbol \theta = (\theta_1,\dots,\theta_{M+1})$
of values that it assumes on those intervals.  Here, $N_e(t) = \theta_k$ for $x_{k-1} \leq t < x_k$,
$k = 1,\dots,M$, where it is understood that $x_0 = 0$.  Also, $N_e(t) = \theta_{M+1}$ for
$t \geq x_M$.  Note that $x_M$ is the time furthest back into the past at which the effective
population size can change.  The values of the
grid points as well as the number $M$ of total grid points are specified beforehand by the user.
A typical way to select the grid points is to decide on a resolution $M$,
let $x_M$ assume the value furthest back in time for which the data are expected to be informative,
and space the remaining grid points evenly between $x_0 =0$ and $x_M$.  Alternatively, as
discussed in the next section, grid points can be selected to align with covariate sampling times
in order to facilitate the modeling of associations between the effective population size
and external covariates.  
%The point furthest back
%in time for which the data are informative corresponds to the time of the MRCA.  In the case
%of a known genealogy, we can set $x_M$ equal to the time of the MRCA.
\par
Suppose we have $m$ known genealogies $g_1,\dots,g_m$ representing the ancestries of samples from
$m$ separate genetic loci with the same effective population size $N_e(t)$.
%Let $\textbf g = (g_1,g_2,...,g_m)$ be a vector of genealogies representing the ancestry of populations
%with the same effective population size $N_e(t)$, where the time $t$ increases into the past.
We assume \textit{a priori} that the genealogies are independent given
$N_e(t)$.  This assumption implies that the genealogies are unlinked which commonly occurs when
researchers select loci from whole genome sequences or when recombination is very likely, such as
between genes in retroviruses. %PL: another interesting situation in which such a model may be useful is when you have sampled the different loci from different individuals (and not necessarily the same number of samples for each locus).
The likelihood of the vector $\textbf g = (g_1,\dots,g_m)$ of genealogies can then be expressed as
the product of likelihoods of individual genealogies:
\begin{equation}
P(\textbf g | \boldsymbol \theta) = \prod_{i=1}^{m} P(g_i|\boldsymbol \theta) .
\end{equation}

%Let $M$ denote the number of points we desire for a fixed-time
%grid, and let $K$ be a positive real cutoff value.  Then the temporal grid points
%$x_1,...,x_M$ are $x_1 = \frac{K}{M}, x_2 = 2\times\frac{K}{M},..., x_M = K$.  Here, we assume the grid points
%are equally spaced, but the model easily extends to arbitrarily spaced grid points.
%\par
%We estimate the effective population size as a piecewise constant function that
%changes values only at grid points.  The cutoff value is the time furthest back into the past at
%which the effective population size changes.
%Notice that for all times
%$t \geq K$ further into the past than the cutoff value, $N_e(t) = N_e(K)$.
%Let $\theta = (\theta_1,...,\theta_{M+1})$ be the
%vector of effective population sizes.  Here, $N_e(t) = \theta_k$ for $x_{k-1} \leq t < x_k$,
%$k = 1,...,M$ where it is understood that $x_0 = 0$.  Also, $N_e(t) = \theta_{M+1}$ for
%$t \geq x_M$.
\par
To construct the likelihood of genealogy $g_i$, let $t_{0_i}$ be the most recent sampling time of sequences contributing to genealogy $i$ and
$t_{\mbox{\tiny MRCA}_i}$ be the time of the MRCA for locus $i$.
%(also referred to as the root height of genealogy $i$).
%PL: suppose you have samples for different individuals in different loci, and hence also different sampling times for different loci, does it require the grid points to be older than the oldest most recent sampling time across all loci or older than at least one of the most recent sampling times?
Let  $x_{\alpha_i}$ denote the minimal grid point greater than
at least one sampling time in the genealogy, and $x_{\beta_i}$ the greatest grid point less than
at least one coalescent time.  Let $u_{ik} = [x_{k-1},x_{k}]$, $k = \alpha_i+1,\dots,\beta_i$,
$u_{i\alpha_i} = [t_{0_i}, x_{\alpha_i}]$, and $u_{i(\beta_i + 1)} = [x_{\beta_i}, t_{\mbox{\tiny MRCA}_i}]$.
For each $u_{ik}$ we let $t_{kj}$, $j=1,\dots,r_k$, denote the ordered times of the grid points and
sampling and coalescent events in the interval.  With each $t_{kj}$ we associate an indicator
$\phi_{kj}$ which takes a value of 1 in the case of a coalescent event and 0 otherwise.  Finally,
let $v_{kj}$ denote the number of lineages present in the genealogy in the interval
$[t_{kj}, t_{k(j+1)}]$.
Following \citet{Griffiths1994}, the likelihood of observing an interval is
\begin{eqnarray}
P(u_{ik}|\theta_k) & = & \prod_{1 \leq j < r_k: \phi_{kj}=1} \frac{v_{kj}(v_{kj}-1)}{2\theta_k}
\prod_{j=1}^{r_k-1} \exp \left[-\frac{v_{kj} (v_{kj}-1)(t_{k(j+1)}-t_{kj})}{2\theta_k} \right],
\end{eqnarray}
for $k = \alpha_i , \dots , \beta_i +1$.

The product of interval likelihoods (5.3) yields the likelihood of coalescent times given the sampling times
with genealogy $g_i$.
To obtain the likelihood of the genealogy, however, we must account for the specific lineages
that merge and result in coalescent events.
Let $P_* (u_{ik}|\theta_k)$ denote $P(u_{ik}|\theta_k)$ except with factors
of the form $\frac{v_{kj}(v_{kj}-1)}{2\theta_k}$ replaced by $\frac{2(2-1)}{2\theta_k} = \frac{1}{\theta_k}$.
Then
\begin{equation}
P(g_i|\boldsymbol \theta) = \prod_{k=\alpha_i}^{{\beta_i}+1} P_* (u_{ik}|\theta_k).
\end{equation}
\par
We introduce some notation that will facilitate the derivation of a Gaussian approximation used to
construct a Markov chain Monte Carlo (MCMC) transition kernel.
If $c_{ik}$ denotes the number of coalescent events which occur during interval $u_{ik}$, we
can write
\begin{equation}
P(g_i| \boldsymbol \theta) = \prod_{k=\alpha_i}^{\beta_i+1} \left(\frac{1}{\theta_k} \right)^{c_{ik}}
\exp \left[-\frac{SS_{ik}}{\theta_k} \right] ,
\end{equation}
where the $SS_{ik}$ are appropriate constants.
Rewriting this expression in terms of $\gamma_k = \log(\theta_k)$, we arrive at
\begin{equation}
P(g_i|\boldsymbol \gamma) = \prod_{k=\alpha_i}^{\beta_i+1} e^{-\gamma_k c_{ik}} \exp[-SS_{ik} e^{-\gamma_k}]
= \prod_{k=\alpha_i}^{\beta_i+1} \exp[-\gamma_k c_{ik} - SS_{ik} e^{-\gamma_k}].
\end{equation}
Invoking conditional independence of genealogies, the likelihood of the vector $\textbf g$ of genealogies is
\begin{eqnarray}
P(\textbf g| \boldsymbol \gamma) & = & \prod_{i=1}^m P(g_i| \boldsymbol \gamma) \\
& = & \prod_{i=1}^m \prod_{k=\alpha_i}^{\beta_i+1} \exp[-\gamma_k c_{ik} - SS_{ik} e^{-\gamma_k}] \\
& = & \exp
\left[ \sum_{k=1}^{M+1} \left[-\gamma_k c_k - SS_k e^{-\gamma_k} \right]  \right]
\end{eqnarray}
where $c_k = \sum_{i=1}^{m} c_{ik}$ and $ SS_k = \sum_{i=1}^{m} SS_{ik}$;
here, $c_{ik}=SS_{ik}=0$ if $k \notin [\alpha_i,\beta_i+1]$.

\par
The Skygrid incorporates the prior assumption that effective population size changes continuously over time
by placing a GMRF prior on $\boldsymbol \gamma$:
\begin{equation}
P(\boldsymbol \gamma|\tau) \propto \tau^{M/2}  \exp \left[-\frac{\tau}{2} \sum_{i=1}^M (\gamma_{i+1}-\gamma_i)^2
\right] .
\end{equation}
This prior does not inform the overall level of the effective population size, just the smoothness of the
trajectory.  One can think of the prior as a first-order unbiased random walk with normal increments.  The precision
parameter $\tau$ determines how much differences between adjacent log effective population size values
are penalized.
%Let $\textbf Q$ be an $(M+1) \times (M+1)$ matrix with
%entries $Q_{ij} = -1$ for $j=i+1$ and $j=i-1$, $Q_{ii} = 2$ for $i = 2,..,M$ and
%$Q_{ii} = 1$ for $i = 1,M+1$.  Then we can write
%\begin{equation}
%P(\gamma|\tau) \propto \tau^{M/2}  \exp \left[-\frac{\tau}{2} \gamma' Q \gamma \right].
%\end{equation}
We assign $\tau$ a gamma prior:
\begin{equation}
P(\tau) \propto \tau^{a-1} e^{-b \tau}.
\end{equation}
In absence of prior knowledge about the smoothness of the effective population size trajectory,
we choose $a = b = 0.001$ so that it is relatively uninformative.
Conditioning on the vector of genealogies, we obtain the posterior distribution
\begin{equation}
P(\boldsymbol \gamma,\tau | \textbf g) \propto P(\textbf g|\boldsymbol \gamma)P(\boldsymbol \gamma|\tau)P(\tau) .
\end{equation}

\subsection{Incorporating Covariates}
\par
We can incorporate covariates into our inference framework by adopting a generalized linear model (GLM) approach.
Let $Z_1,\dots,Z_P$ be a set of $P$ predictors.  Each covariate $Z_j$ is observed or measured at $M+1$ time points,
$t_1,\dots,t_M,t_{M+1}$.  
Here, $t_0=0$ is the most recent sequence sampling time,
$t_i$ denotes the units of time before $t_0$, and $t_0 < t_1 < \dots < t_M < t_{M+1}$.
Alternatively, the covariate may correspond to time intervals
$[t_0,t_1],\dots,[t_{M-1},t_M],[t_M,t_{M+1}]$ rather than time points
(for example, the yearly incidence or prevalence of viral infections).  
In any case,
$Z_{ij}$ denotes covariate $Z_j$ at time point or interval $i$.
Skygrid grid points are chosen to match up with
measurement times (or measurement interval endpoints): $x_1 = t_1,\dots,x_M = t_M$.  Then
$N_e(t) = \theta_{k}$ for $x_{k-1} \leq t \leq x_{k}$, $k = 1,\dots,M$, and
$N_e(t) = \theta_{M+1}$ for $t \geq x_M$.  In our GLM framework, we model the effective population
size on a given interval as a log-linear function of covariates
\begin{equation}
\gamma_{k} =  \log \theta_{k} = \beta_1 Z_{k1} + \dots + \beta_P Z_{kP} + w_k .
\end{equation}
Here, we can impose temporal dependence by modeling $w = (w_1,\dots,w_{M+1})$ as a zero-mean Gaussian process.
Adopting this viewpoint, we propose the following GMRF smoothing prior on $\boldsymbol \gamma$:
\begin{equation}
P(\boldsymbol \gamma| \textbf Z, \boldsymbol \beta,\tau) \propto
\tau^{M/2} \exp \left[-\frac{\tau}{2} (\boldsymbol \gamma - \textbf Z \boldsymbol \beta)' \textbf Q (\boldsymbol \gamma- \textbf Z \boldsymbol \beta) \right].
\end{equation}
In this prior, $\textbf Z$ is an $(M+1) \times P$ matrix of covariates and $\boldsymbol \beta$ is a $P \times 1$ vector
of coefficients representing the effect sizes for the predictors, quantifying their contribution
to $\boldsymbol \gamma$.
Precision $\textbf Q$ is an $(M+1) \times (M+1)$ tri-diagonal matrix with off-diagonal elements equal to $-1$,
$Q_{11}=Q_{M+1,M+1} = 1$,  and $Q_{ii} = 2$ for $i=2,\dots,M$.  Let $\boldsymbol \gamma_{-i}$ denote the
vector obtained by excluding only the $i^{\text{\small th}}$ component from vector $\boldsymbol \gamma$.  Therefore, conditional
on $\boldsymbol \gamma_{-i}$, $\gamma_i$ depends only on its immediate neighbors.
Let $\textbf Z_i$ denote the $i^{\text{\small th}}$ row of covariate matrix $\textbf Z$.
The individual components of $\gamma$ have full conditionals
\begin{eqnarray}
\gamma_1 | \boldsymbol \gamma_{-1} & \sim & N\left(\textbf Z_1' \boldsymbol \beta - \textbf Z_2' \boldsymbol \beta + \gamma_2 , \frac{1}{\tau} \right), \\
\gamma_i | \boldsymbol \gamma_{-i} & \sim & N\left(\textbf Z_i' \boldsymbol \beta + \frac{\gamma_{i-1} + \gamma_{i+1} - \textbf Z_{i-1}' \boldsymbol \beta- \textbf Z_{i+1}' \boldsymbol \beta}{2} , \frac{1}{2\tau} \right) \\
& & \quad \text{for } i=2,\dots,M, \nonumber \\
\gamma_{M+1} | \boldsymbol \gamma_{-(M+1)} & \sim & N \left(\textbf Z_{M+1}' \boldsymbol \beta - \textbf Z_M' \boldsymbol \beta+\gamma_M , \frac{1}{\tau} \right).
\end{eqnarray}
As in the original Skygrid GMRF prior, the precision parameter $\tau$ governs the smoothness of the trajectory and is assigned a gamma prior
\begin{equation}
P(\tau) \propto \tau^{a-1} e^{-b \tau}.
\end{equation}
To complete the model specification, we place a relatively uninformative multivariate normal prior $P(\boldsymbol \beta)$ on
the coefficients $\boldsymbol \beta$.  This yields the posterior
\begin{equation}
P(\boldsymbol \gamma, \boldsymbol \beta, \tau| \textbf g, \textbf Z) \propto
P(\textbf g| \boldsymbol \gamma) P(\boldsymbol \gamma|\textbf Z,\boldsymbol \beta,\tau) P(\boldsymbol \beta) P(\tau) .
\end{equation}

\subsection{Missing Covariate Data}

It is important to have a mechanism for dealing with unobserved covariate values.  This is
particularly crucial because
the population history timeline, which ranges from the most recent sampling time to the time of the
MRCA, necessitates observations from a wide and \textit{a} \textit{priori}
unknown time span.
%The Skygrid enables partitioning of the population history
%timeline into intervals that are maximally compatible with the temporal spacing of the
%covariates.
Let $\textbf Z^{\text{obs}}$ denote the observed covariate values and
$\textbf Z^{\text{mis}}$ the missing covariate values, so that
$\textbf Z = (\textbf Z^{\text{obs}}, \textbf Z^{\text{mis}})$.  The missing data
can be treated as extra unknown parameters in a Bayesian model, and they can be estimated
provided that there is a model that links them to the observed data and other model parameters.
We have the factorization
\begin{equation}
P(\boldsymbol \gamma, \textbf Z^{\text{mis}} |\textbf Z^{\text{obs}}, \boldsymbol \beta, \tau)
= P(\boldsymbol \gamma |\textbf Z^{\text{obs}} ,\textbf Z^{\text{mis}}, \boldsymbol \beta, \tau)
P(\textbf Z^{\text{mis}} |\textbf Z^{\text{obs}}, \boldsymbol \beta, \tau),
\end{equation}
and the marginal density $P(\boldsymbol \gamma |\textbf Z^{\text{obs}}, \boldsymbol \beta, \tau)$
can be recovered by integrating out the missing data.
As a starting point, we assume a missing completely at random structure, meaning that the probability that a
covariate value is missing is independent of observed trait values and other model
parameters.  For the priors on missing covariate values in (21), we can
adopt uniform distributions over plausible ranges.
\par
Alternatively, we can formulate a prior on the missing covariate data that makes use of the
observed covariate values.  Here, we focus on a common scenario where covariate $j$ is observed at times
$x_0, \dots ,x_{K}$ and unobserved at times $x_{K+1}, \dots ,x_{M}$.
Thus, we can write
$\textbf Z^{\text{obs}}_j = (Z_{0j},\dots,Z_{Kj})'$ and $\textbf Z^{\text{mis}}_j = (Z_{(K+1)j},\dots,Z_{Mj})'$.
We model the joint distribution of the observed and missing covariate values as multivariate
normal,
\begin{equation}
\left(
\begin{array}{c}
\textbf Z^{\text{obs}}_j \\
\textbf Z^{\text{mis}}_j \\
\end{array}
\right)
\sim N \left(
\left(
\begin{array}{c}
\boldsymbol \mu_1 \\
\boldsymbol \mu_2 \\
\end{array}
\right) ,
\left(
\begin{array}{c c}
\textbf P_{11} & \textbf P_{12} \\
\textbf P_{21} & \textbf P_{22} \\
\end{array}
\right)^{-1}
\right),
\end{equation}
where
\begin{equation}
\textbf P =
\left(
\begin{array}{c c}
\textbf P_{11} & \textbf P_{12} \\
\textbf P_{21} & \textbf P_{22} \\
\end{array}
\right)
\end{equation}
is the precision matrix.  To impose a correlation structure that enforces
dependence between covariate values corresponding to adjacent times, we
adopt a first-order random walk with full conditionals
\begin{eqnarray}
Z_{0j} | Z_{-0j} & \sim & N\left( Z_{1j} , \frac{1}{\kappa} \right), \\
Z_{ij} | Z_{-ij} & \sim & N\left(\frac{Z_{(i-1)j}+Z_{(i+1)j}}{2}, \frac{1}{2 \kappa} \right) \\
& & \quad \text{for } i=1,\dots,M-1, \nonumber \\
Z_{Mj} | Z_{-Mj} & \sim & N \left( Z_{(M-1)j}, \frac{1}{\kappa} \right).
\end{eqnarray}
Let $\textbf Z^K$ denote a vector of dimension $M-K$ with every entry equal to $Z_{Kj}$.  Then
the distribution of missing covariate values conditional on observed covariate values is
\begin{equation}
P(\textbf Z^{\text{mis}}_j | \textbf Z^{\text{obs}}_j)
\propto \kappa^{(M-K)/2}
\exp \left(-\frac{\kappa}{2}(\textbf Z^{\text{mis}}_j  - \textbf Z^K)'
 \textbf P_{22} (\textbf Z^{\text{mis}}_j  - \textbf Z^K )
\right),
\end{equation}
where
\begin{equation}
\textbf P_{22} =
\left(
\begin{array}{c c c c c}
-1 & 2 & -1 & & \\
& & \ddots & \ddots & \\
& & -1 & 2 & -1 \\
& & & -1 & 1 \\
\end{array}
\right).
\end{equation}

\subsection{Markov Chain Monte Carlo Sampling Scheme}
\par
We use MCMC sampling
to approximate the posterior
\begin{equation}
P(\boldsymbol \gamma, \boldsymbol \beta, \tau| \textbf g, \textbf Z) \propto
P(\textbf g| \boldsymbol \gamma) P(\boldsymbol \gamma|\textbf Z,\boldsymbol \beta,\tau) P(\boldsymbol \beta) P(\tau) .
\end{equation}
To sample $\boldsymbol \gamma$ and $\tau$, we propose
a fast-mixing, block-updating
MCMC sampling scheme for GMRFs \citep{Knorr-Held2002}.
Suppose we have current parameter values $(\boldsymbol \gamma^{(n)}, \tau^{(n)})$.
First, consider the full conditional density
\begin{eqnarray}
P(\boldsymbol \gamma|\textbf g, \textbf Z, \boldsymbol \beta, \tau)
& \propto &  P(\textbf g|\boldsymbol \gamma)
P(\boldsymbol \gamma|\textbf Z, \boldsymbol \beta, \tau) \nonumber \\
& \propto & \exp \left[\sum_{k=1}^{M+1} (-\gamma_k c_k - SS_k e^{-\gamma_k}) \right]
 \tau^{M/2} \exp \left[-\frac{\tau}{2} (\boldsymbol \gamma - \textbf Z \boldsymbol \beta)' \textbf Q
 (\boldsymbol \gamma- \textbf Z \boldsymbol \beta) \right] \nonumber \\
 & = & \tau^{M/2} \exp \left[
 -\frac{\tau}{2} (\boldsymbol \gamma - \textbf Z \boldsymbol \beta)' \textbf Q
 (\boldsymbol \gamma- \textbf Z \boldsymbol \beta)
 -\sum_{k=1}^{M+1} (\gamma_k c_k + SS_k e^{-\gamma_k}) \right] \nonumber \\
 & = & \tau^{M/2} \exp \left[
 -\frac{\tau}{2} \boldsymbol \gamma' \textbf Q \boldsymbol \gamma + (\textbf Z \boldsymbol \beta)' \tau \textbf Q \boldsymbol \gamma
 -\sum_{k=1}^{M+1} (\gamma_k c_k + SS_k e^{-\gamma_k}) \right] .
\end{eqnarray}
Let $h_k (\gamma_k) = (\gamma_k c_k + SS_k e^{-\gamma_k})$.  We can approximate each term
$h_k (\gamma_k)$ by a second-order Taylor expansion
about, say, $\hat{\gamma_k}$:
\begin{eqnarray}
h_k (\gamma_k) & \approx & h_k (\hat{\gamma_k}) + h'_k (\hat{\gamma_k})(\gamma_k - \hat{\gamma_k})
+ \frac{1}{2} h''_k (\hat{\gamma_k})(\gamma_k - \hat{\gamma_k})^2 \nonumber \\
& = & SS_k e^{-\hat{\gamma_k}} \left(\frac{1}{2}\hat{\gamma_k}^2 + \hat{\gamma_k} + 1 \right) \nonumber \\
& & + \left[c_k - SS_k e^{-\hat{\gamma_k}} - SS_k e^{-\hat{\gamma_k}} \hat{\gamma_k}\right] \gamma_k\nonumber \\
& & + \left[\frac{1}{2} SS_k e^{-\hat{\gamma_k}} \right] \gamma^2_k .
\end{eqnarray}
We center the Taylor expansion about a point
$\hat{\boldsymbol \gamma} = (\hat{\gamma}_1, \ldots, \hat{\gamma}_{M+1})$ obtained iteratively by the Newton-Raphson method:
\begin{equation}
\boldsymbol \gamma_{(n+1)} = \boldsymbol \gamma_{(n)} -
[d^2 f(\boldsymbol \gamma_{(n)})]^{-1} (df(\boldsymbol \gamma_{(n)}))'
\end{equation}
with $\boldsymbol \gamma_{(0)} = \boldsymbol \gamma^{(n)}$, the current value of $\boldsymbol \gamma$.
Here,
\begin{equation}
f(\boldsymbol \gamma) = -\frac{1}{2} \boldsymbol \gamma' \tau \textbf Q \boldsymbol \gamma
+ (\textbf Z \boldsymbol \beta)' \tau \textbf Q \boldsymbol \gamma
-\sum_{k=1}^{M+1} (\gamma_k c_k + SS_k e^{-\gamma_k})
\end{equation}
with
\begin{equation}
df(\boldsymbol \gamma) = -\boldsymbol \gamma' \tau \textbf Q
+ (\textbf Z \boldsymbol \beta)' \tau \textbf Q
- [c_1 - SS_1e^{-\gamma_1},...,c_{M+1} - SS_{M+1}e^{-\gamma_{M+1}}]
\end{equation}
and
\begin{equation}
d^2 f(\boldsymbol \gamma) = -\tau \textbf Q - \mbox{diag}[SS_k e^{-\gamma_k}].
\end{equation}
Replacing the terms $h_k(\gamma_k)$ with their Taylor expansions
yields the following second-order Gaussian approximation to the full conditional
density $P(\boldsymbol \gamma|\textbf g, \textbf Z, \boldsymbol \beta, \tau)$ :
%\begin{equation}
%\begin{split}
\begin{multline}
\label{eq:mult}
P(\boldsymbol \gamma|\textbf g, \textbf Z, \boldsymbol \beta, \tau) \approx
\tau^{M/2} \exp\left[-\frac{1}{2} \boldsymbol \gamma' [\tau \textbf Q + \mbox{Diag}(SS_k e^{-\hat{\gamma_k}}) ]
\boldsymbol \gamma + (\tau \textbf Q \textbf Z \boldsymbol \beta)' \boldsymbol \gamma \right. \\
\left.
-\sum_{k=1}^{M+1} (c_k - SS_k e^{-\hat{\gamma_k}} - SS_k e^{-\hat{\gamma_k}} \hat{\gamma_k}) \gamma_k \right],
\end{multline}
%\end{split}
%\end{equation}
where $\mbox{Diag}(\cdot)$ is a diagonal matrix.
\par
Starting from current parameter values $(\boldsymbol \gamma^{(n)}, \tau^{(n)})$,
we first generate a candidate value for the
precision, $\tau^* = \tau^{(n)} f$, where $f$ is drawn from a symmetric proposal distribution with density
$P(f) \propto f + \frac{1}{f}$ defined on $[1/F,F]$.  The tuning constant $F$ controls the distance
between the proposed and current values of the precision.  Next, conditional on $\tau^*$, we propose
a new state $\boldsymbol{\gamma}^*$ using the Gaussian approximation (\ref{eq:mult}) to the full conditional density
$P(\gamma|\textbf g, \textbf Z, \boldsymbol \beta, \tau^*)$.
In the final step, the candidate state $(\tau^*,\boldsymbol \gamma^*)$ is accepted or rejected according
to the Metropolis-Hastings ratio \citep{Metropolis1953, Hastings1970}.

\subsection{Genealogical Uncertainty}
\par
In our development thus far, we have assumed the genealogies $g_1,\dots,g_m$ are known and fixed.  However,
in reality we observe sequence data rather than genealogies.  It is possible to estimate genealogies
beforehand from sequence data and then infer the effective population size from fixed genealogies.
However, this ignores the uncertainty associated with phylogenetic reconstruction.  Alternatively,
we can jointly infer genealogies and population dynamics from sequence data by combining
the estimation procedures into a single Bayesian framework.
\par
We can think of the aligned sequence data
$\textbf Y = (Y_1,\dots,Y_m)$ for the $m$ loci as arising from continuous-time Markov chain (CTMC) models for molecular character
substitution that act along the hidden genealogies.  Each CTMC depends on a vector of
mutational parameters $\Lambda_i$, that include, for example, an overall rate multiplier, relative exchange
rates among characters and across-site variation specifications.  We let
$\boldsymbol \Lambda = (\Lambda_1,\dots,\Lambda_m)$.
We then jointly estimate the genealogies, mutational parameters, precision,
and vector of effective population sizes through their posterior distribution
\begin{equation}
P(\textbf g,\boldsymbol \Lambda,\tau,\boldsymbol \gamma | \textbf Y) \propto
\left[
\prod_{i=1}^{m} P(Y_i|g_i,\Lambda_i)
\right]
P(\boldsymbol \Lambda)P(\textbf g| \boldsymbol \gamma)P(\boldsymbol \gamma|\tau)P(\tau)   .
\end{equation}
Here, the coalescent acts as a prior for the genealogies, and we assume that $\boldsymbol \Lambda$ and $\textbf g$ are
\textit{a priori} independent of each other.
Hierarchical models are however available to share information about $\boldsymbol \Lambda$ among loci without strictly
enforcing that they follow the same evolutionary process \citep{Suchard2003b,Edo-Matas2011}.  We implement our
models in the open-source software program BEAST \citep{Drummond2012}.  The posterior distribution is approximated
through MCMC methods.  We combine our block-updating scheme for $\boldsymbol \gamma$ and $\tau$ with
standard transition kernels available in BEAST to update the other parameters.

\section{Results}

\subsection{Expansion in Epizootic Rabies Virus}

Rabies is a zoonotic disease caused by the rabies virus, and is responsible for over 50,000
human deaths annually.  In 
over 99$\%$ of human cases, the rabies virus is transmitted by dogs.  However,
there are a number of other important rabies reservoirs, such as bats and 
several terrestrial carnivore species, including raccoons \citep{WHORabies}.  Epizootic rabies
among raccoons was first identified in the U.S. in Florida in the 1940s, and the affected
area of the subsequent expansion was limited to the southeastern U.S. 
\citep{Kappus1970}.  A second focus of rabies among raccoons emerged in West Virginia in 
the late 1970s due to the translocation of raccoons incubating rabies from the southeastern
U.S.  The virus spread rapidly along the mid-Atlantic coast and northeastern U.S. 
over the following decades, and is
one of the largest documented outbreaks in the history of wildlife rabies \citep{Childs2000}.  
\par     
\citet{Biek2007} examine the population dynamics of the rabies epizootic among raccoons in
the northeastern U.S. starting in the late 1970s.  In a spatiogenetic
analysis, \citet{Biek2007} compare a coalescent-based Bayesian Skyline estimate \citep{Drummond2005}
of the demographic history to the spatial expansion of the epidemic.  
In a \textit{post hoc} approach, the authors find
very similar temporal dynamics between the effective population size and the 
15-month moving average of the area (in square kilometers)
of counties newly affected by the rabies outbreak each month.  The effective 
population size exhibits stages of moderate and rapid growth, 
as well as plateau periods with little or no growth.
Population expansion coincides with time periods during which 
the virus invades new area at a generally increasing rate.  
On the other hand, the effective population size shows little, if any, growth
during periods when the virus invades new area at a 
declining rate.   
These trends can be seen in Figure 1, which depicts a Skygrid demographic reconstruction
from sequence data along with 
the monthly area newly affected by the virus as a solid black line.
Notably,
\citet{Biek2007} demonstrate through their analysis that the largest contribution to the population expansion
comes from the wave front, highlighting the degree to which the overall viral dynamics 
depend on processes at the wave front.
\par
We build upon the analysis of \citet{Biek2007} by incorporating the spatiotemporal spread of rabies
into the demographic inference model through the Skygrid.  The data consist of 47 sequences
with sampling dates between 1982 and 2004.  As a covariate, we initially adopt the 15-month
moving average of the log-transformed area of all counties newly affected by the raccoon rabies virus each
month from 1977-1999 \citep{Biek2007}.  We infer a posterior mean covariate 
effect size of 0.24 with a 95$\%$ Bayesian credibility interval (BCI) of (-0.77, 1.27), 
implying that there is not a significant association between the log effective population 
size and the covariate.  This is not surprising, considering the patterns of growth and 
decline in the covariate compared with the essentially monotonic trend in the log effective 
population size (see Figure 1).
\par
Graphically comparing the rate at which the virus invades new area with population dynamics
clearly illustrates the relationship between the demographic and spatial expansion of the raccoon
rabies outbreak.  
In modeling the association between the population dynamics and a covariate, however,
we relate the covariate to the total effective population size (as opposed to the change in the effective
population size).  In this case, the cumulative affected area is a 
more suitable covariate than the newly affected area.   
We conduct an additional Skygrid analysis and use the log-transform of
the cumulative area (in square
kilometers) of counties affected by raccoon rabies at various time points
between 1977 and 1999 as a covariate.  The area of a county is added to the cumulative total 
for the month during which rabies is first reported in that county.  There are 175 months
for which the cumulative affected area changes, and we specify the grid points to coincide
with these change points.
%We aim to examine the phenomenon identified by \citet{Biek2007}: that the rate of demographic 
%expansion is largely driven by the wave front. 
%However, we use the cumulative invaded area for the covariate, as opposed to the newly invaded area considered
%by \citet{Biek2007}.   

\begin{figure}
	\centering
	\includegraphics[width=1.0\textwidth]{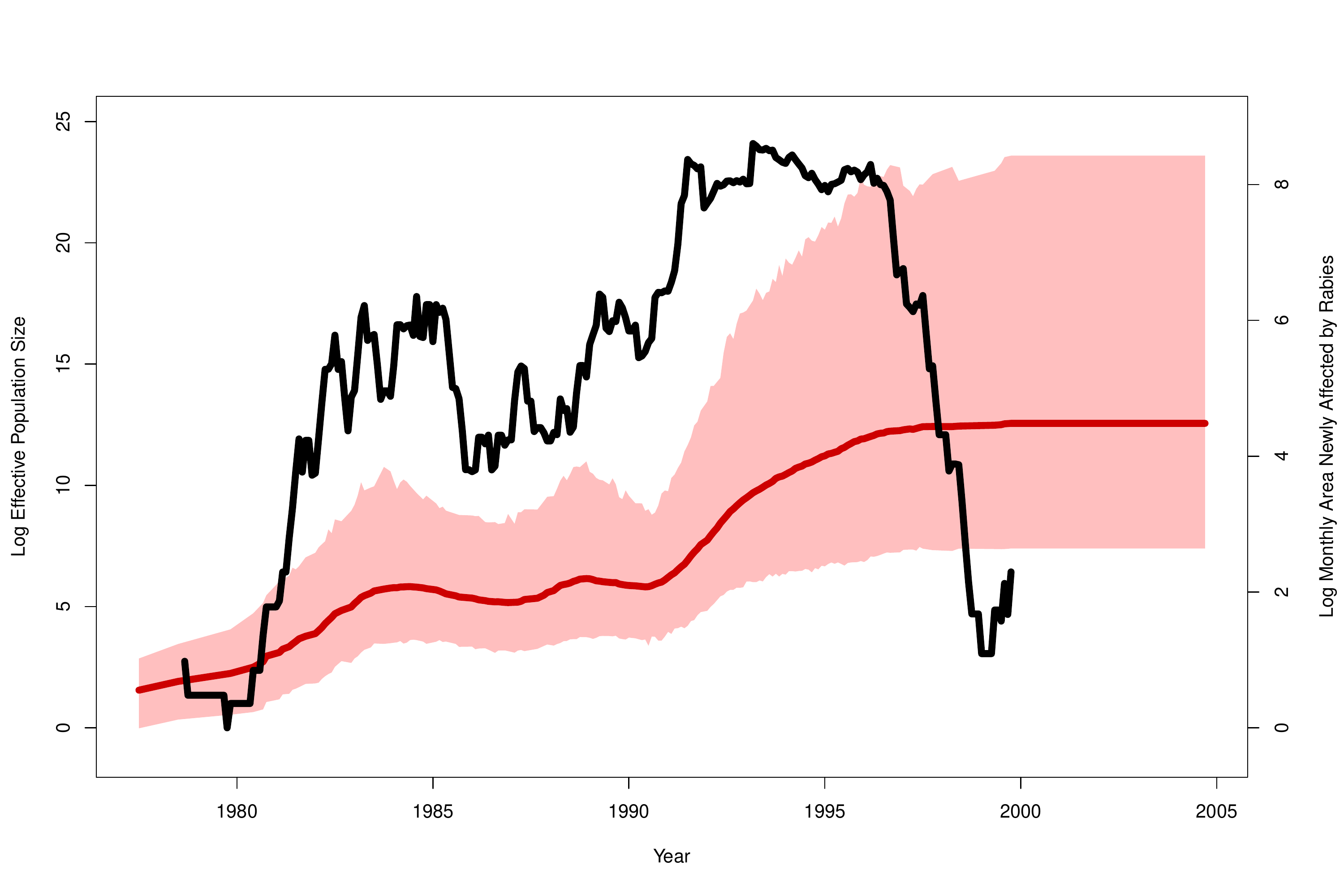}
 	\caption[Demographic history of raccoon rabies epidemic in the northeastern United
 	States]{
 	Demographic history of raccoon rabies epidemic in the northeastern United
 	States.
 	The black line represents the covariate, the
 	15-month moving average of the log-transformed area of all counties newly affected by 
 	the raccoon rabies virus each month.
 	The solid red line is the posterior mean
 	log effective population size trajectory estimated from sequence data.
 	The 95$\%$ Bayesian credibility interval region for the trajectory is shaded in light red.}
 	\label{Fig. 1}
\end{figure}

\begin{figure}
	\centering
	\includegraphics[width=1.0\textwidth]{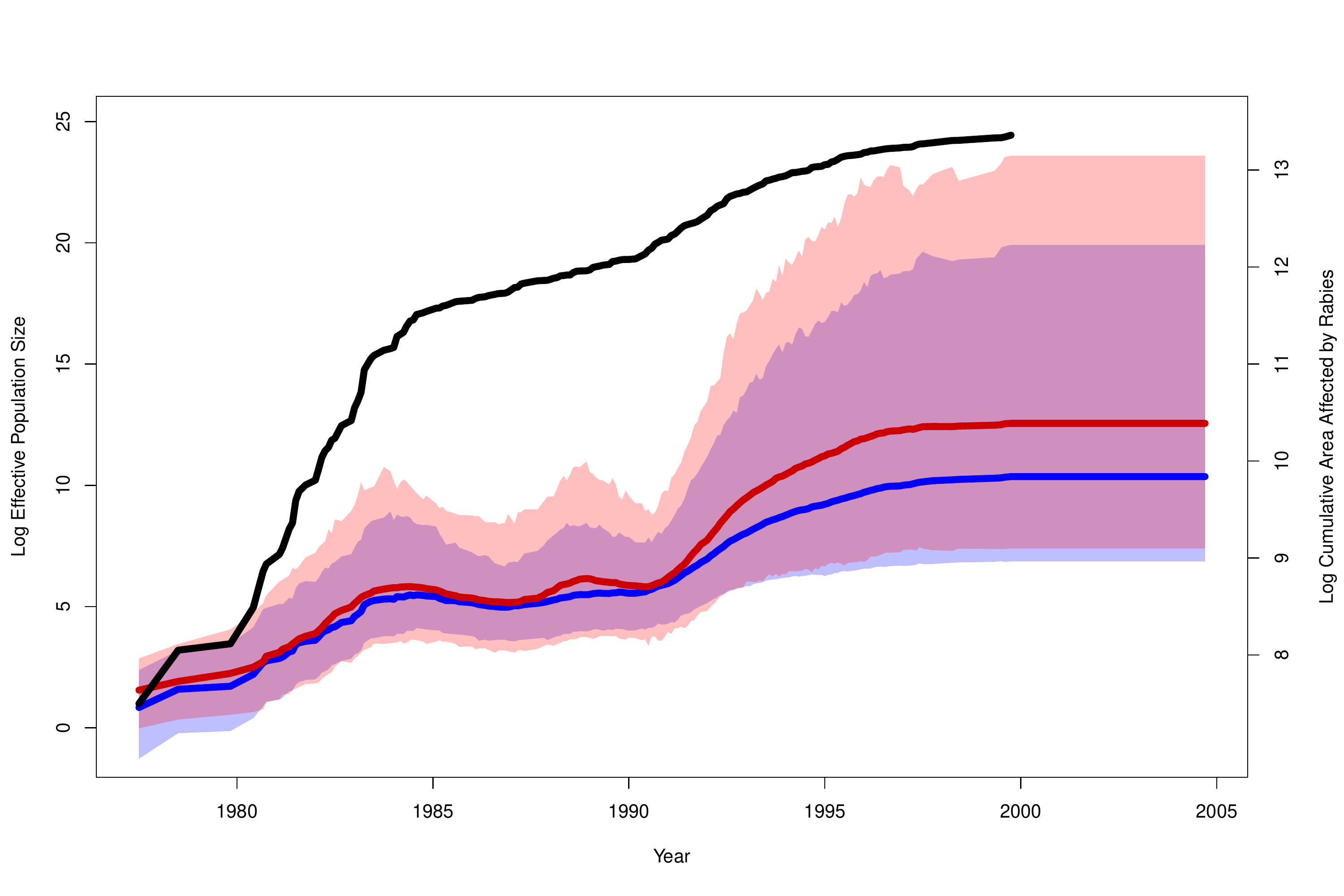}
 	\caption[Demographic history of raccoon rabies epidemic in the northeastern United
 	States]{
 	Demographic history of raccoon rabies epidemic in the northeastern United
 	States.
 	The black line represents the covariate, the log cumulative area of counties affected
 	by raccoon rabies virus.
 	The solid blue line is the posterior mean log effective population size trajectory
 	from the Skygrid analysis with the covariate, and the solid red line is the posterior mean
 	log effective population size trajectory from the Skygrid analysis without the covariate.
 	The 95$\%$ Bayesian credibility interval regions for the trajectories are shaded
 	in light blue for the analysis that includes the covariate and in light red for the
 	analysis without the covariate.
 	The overlap between the two Bayesian credibility interval regions is shaded in purple.}
 	\label{Fig. 2}
\end{figure}

\par
The Skygrid analysis with the log cumulative affected area covariate yields a posterior
mean estimate of 1.30 for the coefficient $\boldsymbol \beta$, with a 95$\%$ BCI
of (0.18, 2.86), implying a significant, positive association between the effective population
size of the raccoon rabies virus and the cumulative area affected by the outbreak.
Skygrid demographic reconstructions of the epidemic are displayed in Figure 2.  The
red line is the posterior mean log effective population size trajectory inferred
using the Skygrid without incorporation of the covariate.  Its 95$\%$ BCI region
is shaded in light red.  The log effective population size trajectory
from the Skygrid analysis with the log cumulative area
covariate is represented by the blue line, and its 95$\%$ BCI region is shaded
in light blue.  The overlap between the two BCI regions is shaded purple.
The log cumulative area affected by the rabies epidemic is depicted as
a solid black line.  Figure 2 shows great correspondence between the temporal dynamics of the demographic and
spatial expansions.  The period 1977-1984 is marked by a steady increase in effective
population size and a rapid exponential increase in affected area.  This is followed in the
period 1984-1990 by a plateau in effective population size along with a much more modest rate
of increase for the affected area.  The affected area begins increasing at a greater rate around
1990 and then plateaus around 1996.  Similary, 1990-1996 marks a stage of demographic expansion,
culminating in a period of stasis from 1996 on.  The two effective population size curves
are nearly identical from 1977-1990 and 1996-2004.  From 1990-1996, both effective population size trajectories
increase, but the rate of increase is more modest in the Skygrid estimate that incorporates the covariate
data.  Notably, the light blue BCI region inferred from the sequence and covariate
data is narrower than and virtually entirely contained within the red BCI region inferred only
from the sequence data.  Thus including the covariate in this analysis not only
yields an estimate consistent with what
we infer from the sequence data alone, but also a more precise estimate.
%\par
%Our results provide further support for the relationship between the demographic and spatial
%patterns identified by \citet{Biek2007}.  When the virus invades new area at an increasing rate,
%the effective population size increases.  When the rate of spatial expansion of the virus decreases,
%the effective population size plateaus.

\subsection{Epidemic Dynamics in Dengue Evolution}

Dengue is a mosquito-borne viral infection that causes a severe flu-like illness in which potentially lethal syndromes occasionally arise.
Dengue is caused by the dengue virus, DENV, an RNA virus which comes in four antigenically distinct but closely
related serotypes, DENV-1 through DENV-4.
\citep{WHODengue}.
A recent estimate places the worldwide burden of dengue at 390 million infections per
year (with 95$\%$ confidence interval 284-528 million), of which 96 million
(67-136 million) manifest clinically (with any level of disease severity) \citep{Bhatt2013}.
%The World Health Organization (WHO)
Dengue is found in tropical and sub-tropical
climates throughout the world, mostly in urban and semi-urban areas \citep{WHODengue}.
\par
Dengue incidence records show patterns of periodicity with
outbreaks every 3-5 years \citep{Cummings2004, Adams2006, Bennett2010}.
%(Talk about explanations of periodicity and get to how it reflects effective population
%size).
Studies have shown that the epidemiological dynamics of dengue transmission in
Puerto Rico are reflective of changes in the viral effective population size
\citep{Bennett2010, Carrington2005}.
\citet{Bennett2010} explore the dynamics of DENV-4 in Puerto Rico from 1981-1998.
By \textit{post hoc} comparing dengue isolate counts to effective population size estimates obtained using the Skyride
model \citep{Minin2008}, \citet{Bennett2010} show that the pattern of cyclic epidemics is highly
correlated with similar fluctuations in genetic diversity.  We build upon their analysis by
inferring the effective population size of DENV-4 in Puerto Rico with DENV-4 isolate counts as
a covariate.
\par
We analyze a data set of 75 DENV-4 sequences, compiled by \citet{Bennett2003} through
sequencing randomly selected DENV-4 isolates from
Puerto Rico from the U.S. Centers for Disease Control
and Prevention (CDC) sample bank.  The sampling dates include 

\begin{figure}
	\centering
	\includegraphics[width=1.15\textwidth]{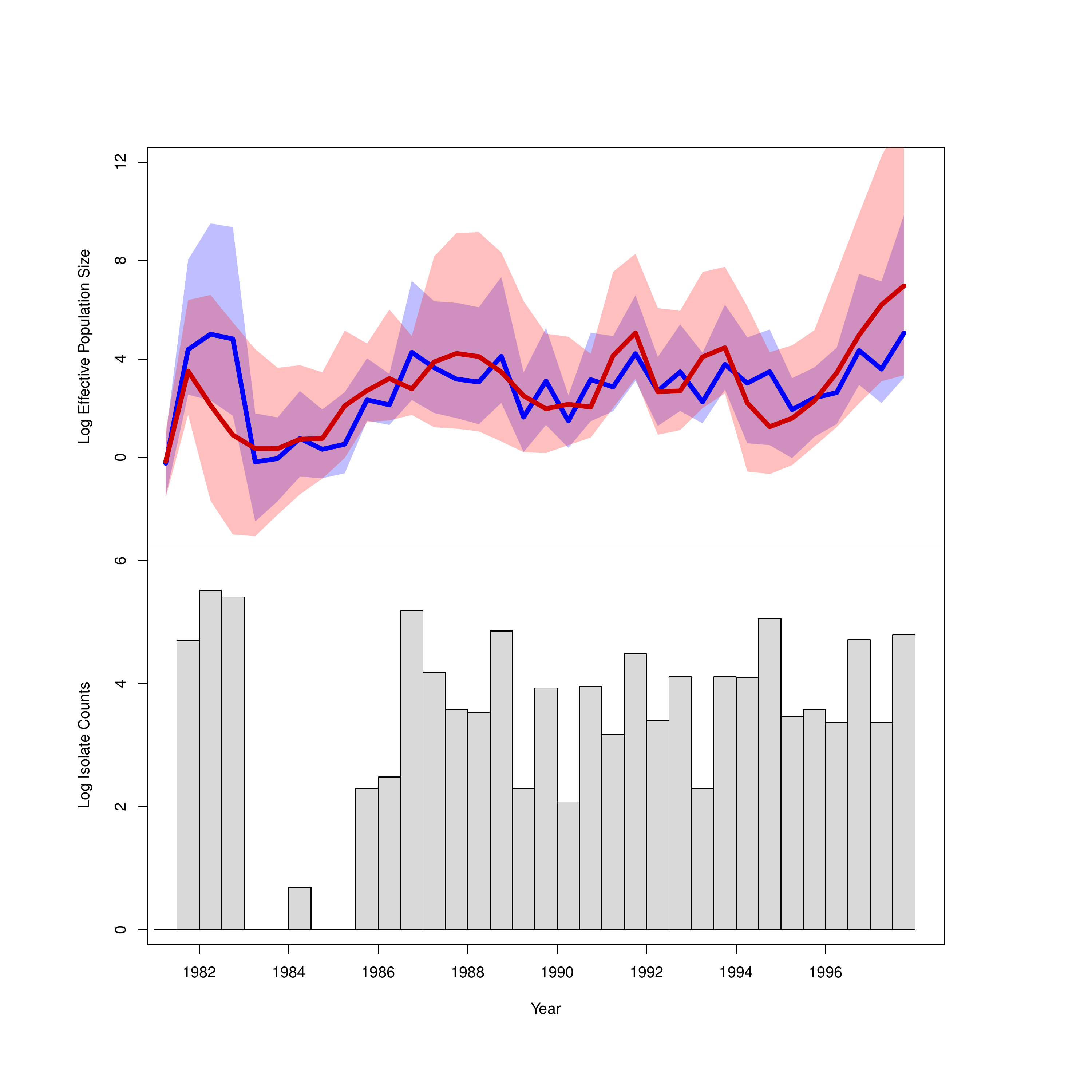}
 	\caption[Population and epidemiological dynamics of DENV-4 virus in Puerto Rico]
 	{Population and epidemiological dynamics of DENV-4 virus in Puerto Rico.
 	The top plot
 	depicts Skygrid effective population size estimates.  See Figure 2 for the legend explanation.
 	%In the top plot,
 	%the solid red line is the posterior mean log effective population size trajectory
 	%from the Skygrid analysis with covariates, and the solid blue line is the same
 	%trajectory from the Skygrid analysis without covariates.
 	%The 95$\%$ Bayesian credibility interval regions for the trajectories are shaded
 	%in light red for the analysis that includes covariates and in light blue for the
 	%analysis without covariates.
 	%The overlap between the two regions is shaded in purple.
 	The bars in the bottom plot represent DENV-4 isolate count covariate data.}
 	\label{Fig. 3}
\end{figure}

\noindent 1982 ($n=14$), 1986/1987 ($n=19$),
1992 ($n=15$), 1994 ($n=14$), and 1998 ($n=13$).  The covariate data consist of the number of
DENV-4 isolates recorded over every 6-month period from 1981-1998.
%talk about isolates vs cases
DENV-4 isolate counts are transformed via the map $x \mapsto \log(x+1)$ (this specific logarithmic
transformation is chosen to accommodate the transformation of isolate counts of zero).
\par
Figure 3 presents the demographic and epidemiological patterns of DENV-4, with the top plot
consisting of Skygrid effective population size estimates and the bottom showing a bar
graph of transformed DENV-4 isolate counts.  The blue and red curves in the top plot
correspond to the posterior mean
log effective population size trajectories from Skygrid analyses with and without the
DENV-4 isolate count covariate, respectively.
The 95$\%$ BCI regions for the trajectories are shaded in light blue for the analysis that
includes the covariate and in light red for the analysis without the covariate.
The overlap between the two BCI regions is shaded in purple.  The demographic
patterns are generally consistent with the isolate count fluctuations, and suggest
a periodicity of 3-5 years.  This concordance
is supported by a positive, statistically significant estimate of the coefficient $\boldsymbol \beta$
relating the effective population size to isolate counts: a posterior mean of 0.90 with
95$\%$ BCI (0.36, 1.69).
\par
While the two effective population size trajectories are similar, they do have some
notable differences.
The blue-colored trajectory inferred from both sequence and covariate
data closely reflects the DENV-4 isolate count patterns, but
the red-colored trajectory inferred entirely from sequence data
diverges from the isolate count trends during certain periods.
First, the red trajectory shows a dramatic increase in effective population size in 1981, consistent
with a rise in DENV-4 isolates.  However, the red trajectory decreases during 1982 while
the DENV-4 isolate counts remain at a high level.
Second, the period from late 1986 to late 1988 begins and ends with relative peaks
in DENV-4 isolates, with a trough
in between.  By contrast, the red curve reaches a peak during the isolate trough and is on the
decline during the late 1988 peak.  Third, the red trajectory shows a trough in the
effective population size during 1994 that
occurs about a year before a similar trough in DENV-4 isolates.  These discrepancies
may be due to biased sampling in isolate counts and reflect limitations of
epidemiological surveillance.  Isolate counts are a rough measure of incidence, and their
error rates are subject to accurate diagnostic rates by medical personnel, reporting rates,
and the rate at which suspected cases are submitted for isolation \citep{Bennett2010}.
On the other hand, the epidemiological trends
are not necessarily incompatible with the effective population size trajectory estimated
entirely from sequence
data when the latter's uncertainty is taken into account.  The blue-colored
trajectory inferred from both sequence and isolate count data
does not deviate from the isolate count data in the ways that the red trajectory does.
However, the blue trajectory lies entirely inside the red 95$\%$ BCI region.
Furthermore, apart from a 1.5 year period in 1981-82,
the blue 95$\%$ BCI region
is virtually entirely contained within, and is narrower than, the red 95$\%$ BCI region.
Therefore, the Skygrid
estimate that incorporates the DENV-4 isolate count covariate yields a demographic pattern
that reflects epidemiological dynamics, and is more precise than, but not incompatible
with, the effective population size estimate inferred only from sequence data.

\subsection{Demographic History of the HIV-1 CRF02\_AG Clade in Cameroon}

Circulating recombinant forms (CRFs) are genomes that result from recombination of two
or more different HIV-1 subtypes and that have been found in at least three epidemiologically
unrelated individuals.  %cite lanl
Although CRF02\_AG is globally responsible for only
7.7$\%$ of HIV infections \citep{Hemelaar:2011fk}, it accounts for 60-70$\%$ of infections
in Cameroon \citep{Brennan2008, Powell2010}.
\par
We investigate the population history of the CRF02\_AG clade in Cameroon by examining
a multilocus alignment of 336 \textit{gag}, \textit{pol}, and \textit{env} CRF02\_AG
gene sequences sampled between 1996 and 2004 from blood donors from Yaounde and Douala \citep{Brennan2008}.
\citet{Faria2011} infer the effective population size from this data set
with a parametric piecewise logistic growth-constant demographic
model.  Their results point to a period of exponential growth up until the mid 1990s, at which
point the effective population size plateaus.  \citet{Gill2013} follow up with a
nonparameteric Skygrid analyis that reveals a monotonic growth in effective population size
that peaks around 1997 and is then followed
by a decline (rather than a plateau) that persists up until the most recent sampling time.
We build upon these analyses by introducing two covariates: the yearly prevalence of HIV in
Cameroon among adults ages 18-49, and the yearly HIV incidence rate in Cameroon among
adults ages 18-49 \citep{UNAIDS}.  UNAIDS prevalence and incidence estimates for Cameroon
only go back to 1990, so we integrate out
the missing covariate values as described in Section 2.4 by modeling the covariate values as a first-order
random walk.
\par
Figure 4 depicts the effective population size trajectory along with
the HIV prevalence data.  
%In contrast to the incidence rate, 
The prevalence increases up until
2000, stays constant for 4 years, and then declines slightly in 2004.  The temporal pattern
of the prevalence differs markedly from that of the demographic history, and this discordance
is reflected in the GLM coefficient quantifying the prevalence effect size.  The coefficient has a posterior mean
of 0.85 with 95$\%$ BCI (-0.18, 2.03), indicating no significant association between the effective
population size and prevalence.
\par
The coefficient quantifying the effect size for the incidence rate covariate has a posterior mean
of 9.20 with 95$\%$ BCI (1.43, 16.17), implying a significant association between the
population history of the CRF02\_AG clade and the HIV incidence rate among adults ages 18-49 in
Cameroon.  As shown in Figure 5, the effective population size and incidence rate display
similar dynamics: both increase up until a peak around 1997, then decline.
The posterior mean log effective population size
and 95$\%$ BCI under the Skygrid model without covariates are
virtually the same as the Skygrid estimates that incorporate the incidence data.
This is in contrast to the
previous examples we've seen, where inclusion of covariates affects effective
population size estimates, and it may reflect the larger amount of sequence data relative
to covariate data in this example.
% need summary sentence on how, in this case, effective population size is more
% reflective of incidence than prevalence
It is notable that in this example the effective population size is more reflective
of incidence than prevalence.  This is in accordance with expectations put forth
by recent epidemiological modeling of infectious disease dynamics \citep{Volz2009, Frost2010}.

\begin{figure}
	\centering
	\includegraphics[width=1.0\textwidth]{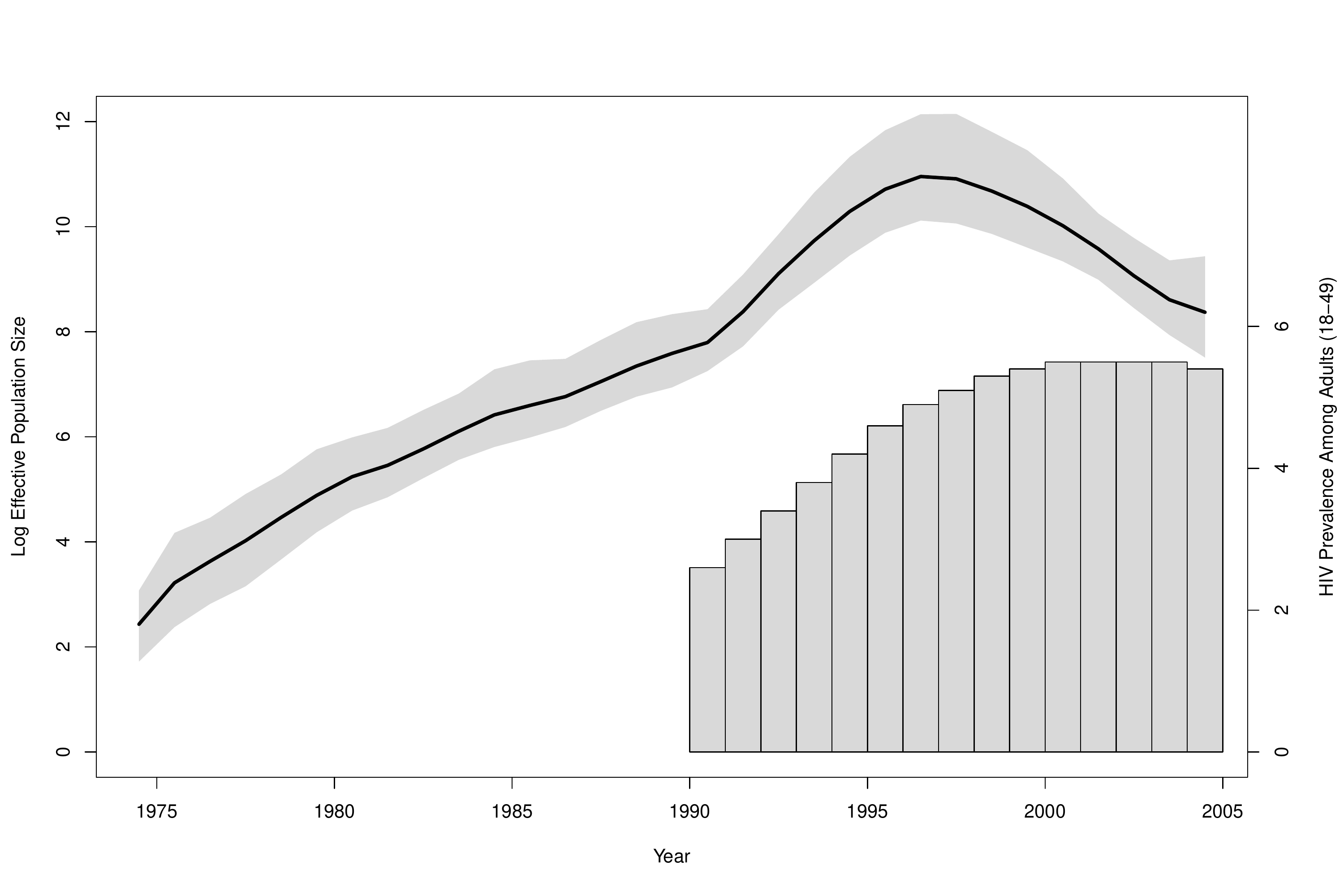}
 	\caption[Demographic history of HIV-1 CRF02\_AG clade in Cameroon with prevalence data.]
 	{Demographic history of HIV-1 CRF02\_AG clade in Cameroon.
 	The solid black line is the posterior mean log effective population size trajectory,
 	and its 95$\%$ Bayesian credibility interval region is shaded in gray.
 	The bars represent HIV prevalence estimates for adults of ages 18-49 in Cameroon.
 	}
 	\label{Fig. 4}
\end{figure}

\begin{figure}
	\centering
	\includegraphics[width=1.0\textwidth]{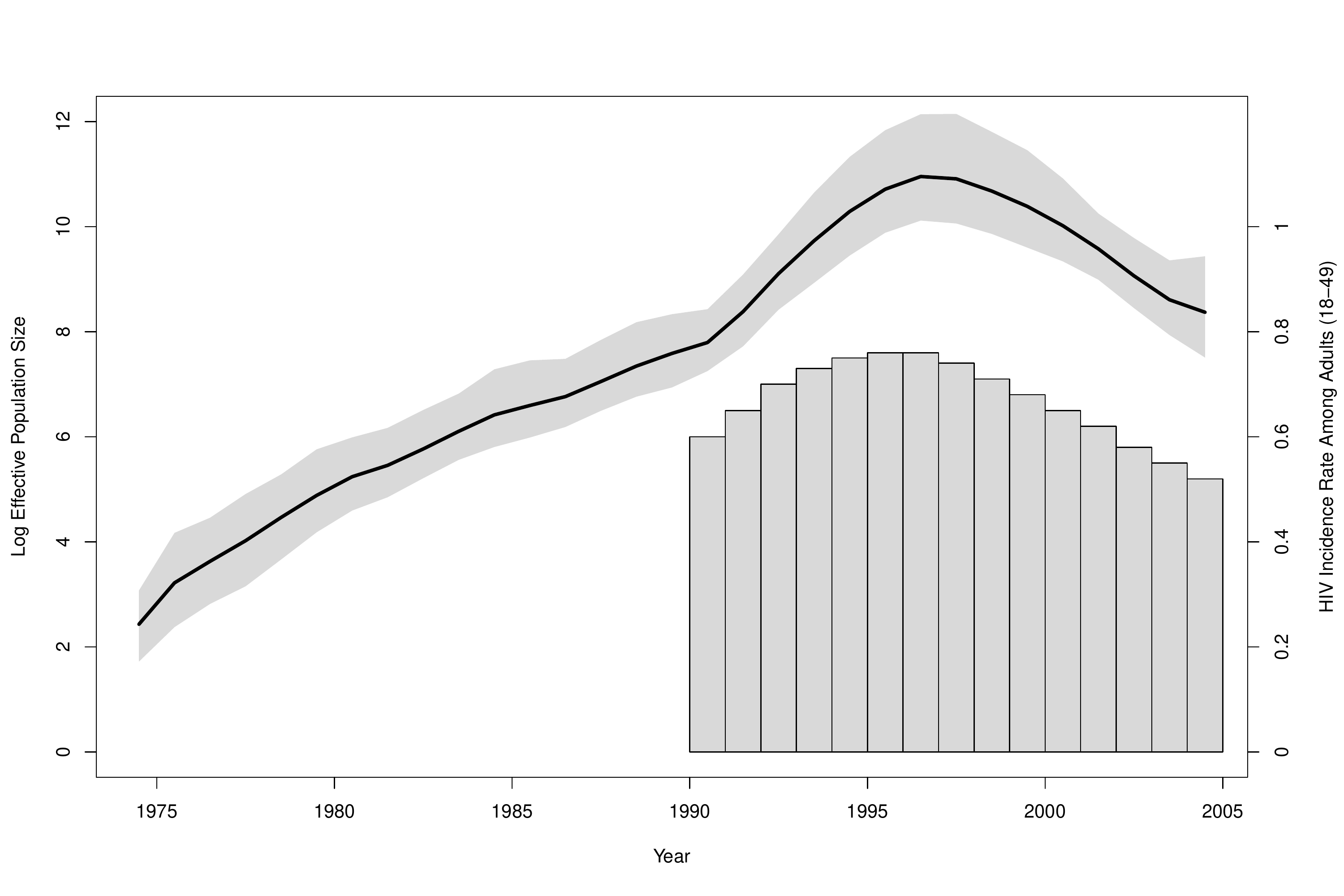}
 	\caption[Demographic history of HIV-1 CRF02\_AG clade in Cameroon with incidence data.]
 	{Demographic history of HIV-1 CRF02\_AG clade in Cameroon.
 	The solid black line is the posterior mean log effective population size trajectory,
 	and its 95$\%$ Bayesian credibility interval region is shaded in gray.
 	The bars represent HIV incidence rate estimates for adults of ages 18-49 in Cameroon.
 	}
 	\label{Fig. 5}
\end{figure}

\subsection{Population Dynamics of Late Quaternary Musk Ox}

Population decline and extinction of
large-bodied mammals characterizes the Late Quaternary period \citep{Barnosky2004, Lorenzen2011}.  The causes of these
megafaunal extinctions remain poorly understood,
and much of the debate revolves around the impact of climate change and humans
\citep{Stuart2004, Lorenzen2011}.  Demographic reconstructions from ancient DNA
enable clarification of the roles of climatic and anthropogenic factors
by providing a means to compare demographic patterns
over geologically significant time scales
with paleoclimatic and fossil records
%records and evidence of human presence in relevant
%times and locations
\citep{Shapiro2004, Lorenzen2011}.
\par
\citet{Campos2010} employ the
Skyride \citep{Minin2008} and Bayesian Skyline \citep{Drummond2005} models to reconstruct
the population dynamics
of musk ox dating back to the late Pleistocene era from ancient DNA sequences.  The musk ox population was
once widely distributed in the holarctic ecozone but is now confined to Greenland and the Arctic Archipelago,
and \citet{Campos2010} explore potential causes of musk ox population decline.  The authors find that the
arrival of humans into relevant areas did not correspond to changes in musk ox effective population size.
On the other hand, \citet{Campos2010} observe that time intervals during which musk ox populations increase generally
correspond to periods of global climatic cooling, and musk ox populations decline during warmer and
climatically unstable periods.  Thus environmental change, as opposed to human presence, emerges as
a more promising candidate as a driving force behind musk ox population dynamics.
\par
We apply our extended Skygrid model
to assess the relationship between the population history of musk ox and climate change.
Oxygen isotope records serve as useful proxies for temperature in ancient climate studies.
Here, we use ice core $\delta^{18}$O data from the Greenland Ice Core Project (GRIP) \citep{Johnsen1997, Dansgaard1993, GRIP1993, Grootes1993, Dansgaard1989}.
$\delta^{18}$O is a measure of oxygen isotope composition.  In the context of ice core data,
lower $\delta^{18}$O values correspond to colder polar temperatures.
As a covariate, we adopt a mean $\delta^{18}$O value, taking the average of
$\delta^{18}$O values corresponding to each 3,000-year interval.  The sequence data consist of
682 bp of the mitochondrial control region, obtained from 149 radiocarbon dated
specimens \citep{Campos2010}.  The ages of the specimens range from the present to 56,900 radiocarbon
(${}^{14}$C) years before present (YBP).  The sampling locations span the demographic
range of ancient musk ox, with samples from the Taimyr Peninsula ($n=54$), the Urals ($n=26$),
Northeast Siberia ($n=12$), North America ($n=14$) and Greenland ($n=43$).
\par
Figure 6 presents the posterior mean log effective population size trajectory (blue line)
along with its 95$\%$ BCI region shaded in light blue.  The $\delta^{18}$O covariate
values are represented by the red line. The plot shows
a steady increase in effective population size up until about 60,000 YBP.  During this period,
the covariate pattern suggests a general trend of cooling, although there is considerable fluctuation
in the mean $\delta^{18}$O values.  The effective population size
plateaus from 60,000 to 55,000 YBP, then decreases from 55,000 to 40,000 YBP.  During the
aforementioned period of demographic decline, the covariate does not display any clear
trends, fluctuating back and forth.  The effective population size increases from about
40,000 to 25,000 YBP while the $\delta^{18}$O covariate continues to fluctuate, although it undergoes a net

\begin{figure}
	\centering
	\includegraphics[width=1.0\textwidth]{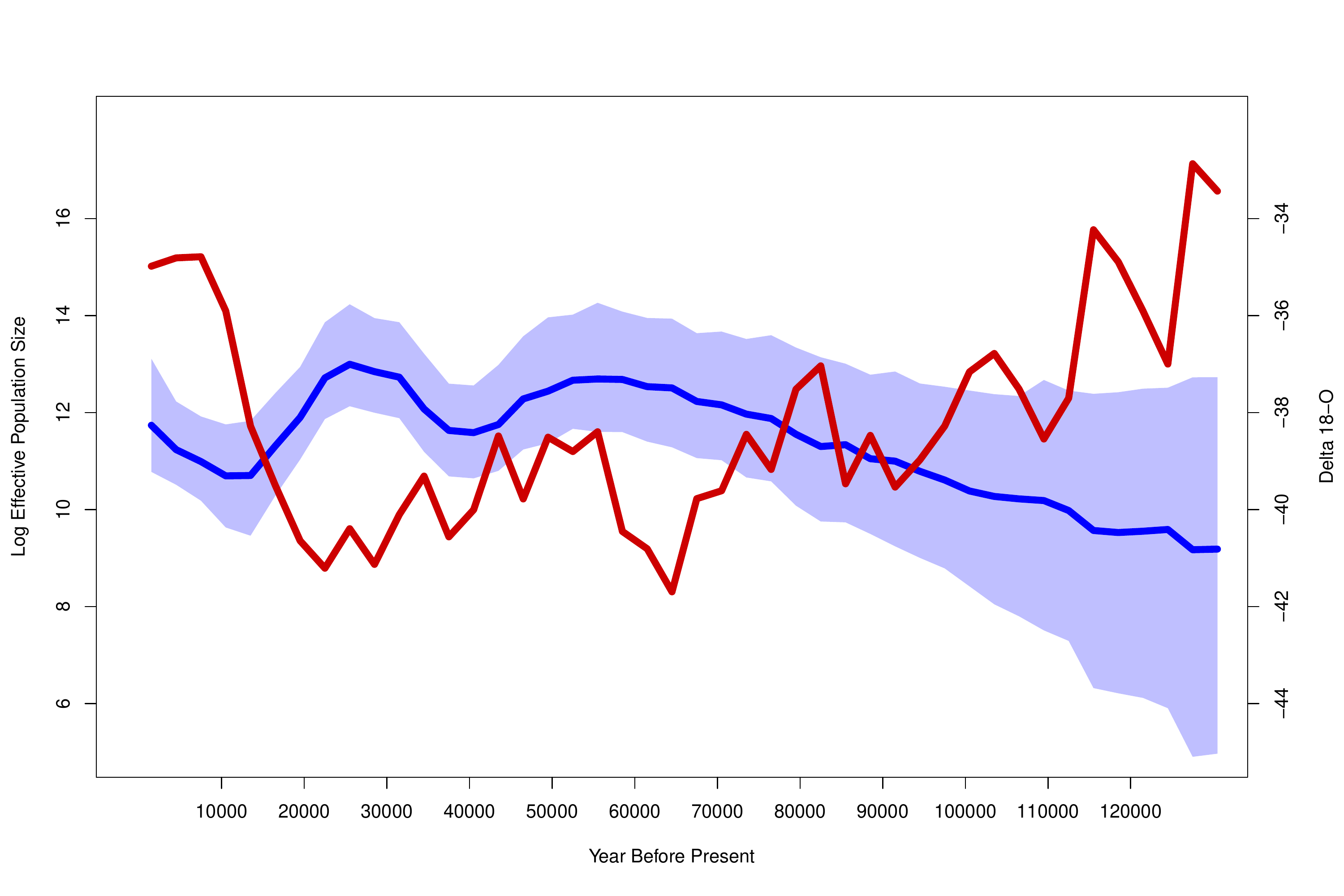}
 	\caption[Demographic history of ancient musk ox]
 	{Demographic history of ancient musk ox.
 	The axis is labelled according
 	to years before present.
 	The solid blue line is the posterior mean log effective population size trajectory,
 	and its 95$\%$ Bayesian credibility interval region is shaded in light blue.
 	The solid red line represents the $\delta^{18}$O covariate.  We do not infer a significant relationship
 	between the effective population size and the covariate.
 	}
 	\label{Fig. 6}
\end{figure}

\noindent decrease from the beginning to the end of the phase.  Notably, the musk ox population reaches its
peak diversity at around
25,000 YBP, coinciding with the Last Glacial Maximum.
The period from 25,000 to 12,000 YBP is marked by a decrease in effective
population size along with a postglacial warming.  
Finally, a demographic recovery and a very mild decline in the $\delta^{18}$O covariate 
characterize the last 12,000 YBP.  The patterns
we note in Figure 6 are consistent with the observations of \citet{Campos2010}.
However, the GLM coefficient $\boldsymbol \beta $ has a posterior mean of
-0.09 with a 95$\%$ BCI of (-0.50, 0.35), indicating that there is not a significant association
between the log effective population size and the $\delta^{18}$O covariate.
This is not surprising upon further reflection.  While the net changes in the covariate
from the beginning to the end of the various monotonic phases of the population trajectory lend
some support to the hypothesis of a negative relationship between the effective population size
and $\delta^{18}$O covariate,
the pervasive covariate fluctuations render the relationship insignificant. 
\par
There are more than 5,000 $\delta^{18}$O measurements in the GRIP data corresponding
to different time points in the musk ox population history timeline.  Our default
approach is to specify Skygrid grid points so that the trajectory has as many piecewise constant
segments as there are covariate measurement times.  To avoid having an inappropriately large
number of change points, however, we've 
used the average of $\delta^{18}$O values corresponding to each 3,000-year interval in the
timeline as a covariate.  Notably, adopting averages over intervals of 
lengths 1,000, 5,000, or 10,000 years as covariates yields the same basic outcome: the effect size of
the covariate is not statistically significant.  
\par
While we do not infer a significant association between the log effective population size and
$\delta^{18}$O covariate values, this does not rule out climate change as a driving force
behind musk ox
population dynamics.  The musk ox is known to be very sensitive to temperature and is not
able to tolerate high summer temperatures \citep{Tener1965}.  Using species distribution models,
dated fossil remains and paleoclimatic data, \citet{Lorenzen2011}
demonstrate a positive correlation between musk ox genetic diversity and its climate-driven
range size over the last 50,000 years.
The $\delta^{18}$O data we use here
do not account for geographic variability in
temperature.
Furthermore, we have not controlled for any potential confounders, such as range size or
proportion of range overlap with humans.
Nevertheless, our analysis serves as a precaution against oversimplification
in the search for explanations of megafaunal population decline and extinctions.
Incorporating additional covariate data into future studies may reveal a more complete, nuanced story of large mammal population dynamics during the
Late Quaternary period.
%It is hoped that additional covariate data can be incorporated into future studies
%to reveal a more complete, nuanced story of large mammal population dynamics during the
%Late Quaternary period.
% address how climate change alone explains the extinction of eurasian musk ox, as asserted by lorenzen et al

\section{Discussion}

We present a novel coalescent-based Bayesian framework for estimation of effective population
size dynamics from molecular sequence data and external covariates.  We achieve this
by extending the popular Skygrid model to incorporate covariates.  In doing so, we retain
the key elements of the Skygrid: a flexible, nonparametric demographic model, smoothing
of the trajectory via a GMRF prior, and accommodation of sequence data from multiple
genetic loci.
\par
Effective population size is of fundamental interest in population genetics, infectious
disease epidemiology, and conservation biology.  It is crucial to identify explanatory factors,
and to achieve a greater understanding of the association between the effective population size and
such factors.  In the context of viruses, it is important to assess the relationship
between effective population size and epidemiological dynamics characterizing the number
of infections and the spatiotemporal spread of an outbreak.  Our extended Skygrid framework
enables formal testing and characterization of such associations.
\par
We showcase our methodology in four examples.
%We infer past population
%dynamics of ancient horses with global surface air temperature as a covariate and find
%a positive, significant association between horse effective population size and temperature.
Our analysis of the raccoon rabies epidemic in the northeastern United States uncovers striking similarities
between the viral demographic expansion and the amount of area affected by the outbreak.
We reconstruct a cyclic pattern for the effective population size of DENV-4 in Puerto Rico,
coinciding with trends in viral isolate count data.  Comparing the population history of the HIV-1
CRF02\_AG clade in Cameroon with HIV incidence and prevalence data reveals a greater alignment
with the HIV incidence rate than the prevalence rate.  Finally, we consider the role of climate change
in ancient musk ox population dynamics by using oxygen isotope data from the GRIP ice core as a proxy
for temperature.  We do not find a significant association, but our analysis demonstrates the need
for a more thorough examination with additional covariates to follow up on previous investigations
of the causes of ancient megafaunal population dynamics that consider a number of different factors.
\par
Simultaneous inference of the effective population size and its association with covariates enables
the uncertainty of the effective population size to be taken into account when assessing the
association.  Post hoc analyses comparing the mean effective population size trajectory with
covariates (employing a standard linear regression approach, for example) are possible.  However,
such approaches may erroneously rule out significant associations by overemphasizing incompatibilities
between the covariates and mean population trajectory.  Furthermore, in the case of
significant associations, regression coefficient estimates that disregard demographic uncertainty may have
inflated precision.
\par
Integrating covariates into the demographic inference framework not only enables testing and
quantification of associations with the effective population size, it also
provides additional information about past population dynamics.  In two of our four
examples, effective population size trajectories inferred from both sequence and covariate data
differ markedly from trajectories inferred only from sequence data.  In the rabies and dengue
examples, the estimates based on sequence and covariate data are essentially consistent with
with the estimates from the sequence data (in terms of the former having BCI regions almost entirely
contained in the BCI regions of the latter), but more precise and more reflective of covariate trends.
%In summary, inferring effective population size from covariates in addition to sequence data can yield more
%precise estimates with dynamics that reflect the additional information provided by covariates while
%still conforming to the specifications suggested by the sequence data.
\par
Our extension of the Skygrid represents a first step toward a more complete understanding of
past population dynamics, and the utility of the approach as demonstrated in the real data examples
is promising.  Our examples have only involved one or two covariates, but our implementation
can support a large number of predictors.  
Furthermore, we plan to equip the Skygrid with efficient variable
selection procedures to identify optimal subsets of predictors
\citep{George1993, Kuo1998, Chipman2001}.  
%We anticipate that analyses with a large number
%of covariates may yield unexpected insights and provide a more comprehensive   
There is considerable potential for further development.  For example, there is a prominent
correspondence between spatial distribution and genetic diversity in the raccoon rabies example,
and in previous studies of megafauna species \citep{Lorenzen2011}.  We envision combining
the Skygrid with phylogeographic inference models \citep{bloomquist2010three} to simultaneously
infer the wave-front of a population from sampling location data and use the wave-front as a predictor to model
the effective population size. 
Attempts to infer associations between covariates and effective population size dynamics can be hampered
by a scarcity of covariate data. Fortunately, there may exist measurements of the same covariates
corresponding to different, but similar, genetic sequence data sets.
We may, for example, have
drug treatment data corresponding to several different HIV patients and wish to assess the
relationship between the drug and intrahost HIV evolution.  In such a setting, Bayesian
hierarchical modeling could enable pooling of information from multiple data sets.
%say something about incorporating covariates in less restrictive coalescent models

\section*{Acknowledgments}
%We would like to thank $\dots$
The research leading to these results has received funding from the European Research Council 
under the European Community's Seventh Framework Programme (FP7/2007-2013) under 
Grant Agreement no.~278433-PREDEMICS and ERC Grant agreement no.~260864 and the 
National Institutes of Health (R01 AI107034, R01 HG006139, R01 LM011827 and 5T32AI007370-24) and the 
National Science Foundation (DMS 1264153).
R.B. was supported by NIH grant RO1 AI047498 and the RAPIDD programme of the Science and 
Technology Directorate of the Department of Homeland Security and NIH Fogarty International Centre.

\clearpage

\bibliographystyle{sysbio}

\bibliography{list_of_references3}

\begin{thebibliography}{69}
\providecommand{\natexlab}[1]{#1}
\providecommand{\selectlanguage}[1]{\relax}
\providecommand{\bibAnnoteFile}[1]{%
  \IfFileExists{#1}{\begin{quotation}\noindent\textsc{Key:} #1\\
  \textsc{Annotation:}\ \input{#1}\end{quotation}}{}}
\providecommand{\bibAnnote}[2]{%
  \begin{quotation}\noindent\textsc{Key:} #1\\
  \textsc{Annotation:}\ #2\end{quotation}}

\bibitem[{Adams et~al.(2006)Adams, Holmes, Zhang, Mammen, Nimmannitya,
  Kalayanarooj, and Boots}]{Adams2006}
Adams, B., E.~Holmes, C.~Zhang, M.~Mammen, S.~Nimmannitya, S.~Kalayanarooj, and
  M.~Boots. 2006. Cross-protective immunity can account for the alternating
  epidemic pattern of dengue virus serotypes circulating in {B}angkok.
  Proceedings of the National Academy of Sciences 103:14234--14239.
\bibAnnoteFile{Adams2006}

\bibitem[{Atkinson et~al.(2008)Atkinson, Gray, and Drummond}]{Atkinson2008}
Atkinson, Q., R.~Gray, and A.~Drummond. 2008. mt{DNA} variation predicts
  population size in humans and reveals a major {S}outhern {A}sian chapter in
  human prehistory. Molecular Biology and Evolution 25:468--474.
\bibAnnoteFile{Atkinson2008}

\bibitem[{Barnosky et~al.(2004)Barnosky, Koch, Feranec, Wing, and
  Shabel}]{Barnosky2004}
Barnosky, A., P.~Koch, R.~Feranec, S.~Wing, and A.~Shabel. 2004. Assessing the
  causes of {L}ate {P}leistocene extinctions on the continents. Science
  306:70--75.
\bibAnnoteFile{Barnosky2004}

\bibitem[{Bazin et~al.(2006)Bazin, Glemin, and Galtier}]{Bazin2006}
Bazin, E., S.~Glemin, and N.~Galtier. 2006. Population size does not influence
  mitochondrial genetic diversity in animals. Science 312:570--572.
\bibAnnoteFile{Bazin2006}

\bibitem[{Bennett et~al.(2010)Bennett, Drummond, Kapan, Suchard, Munoz-Jordan,
  Pybus, Holmes, and Gubler}]{Bennett2010}
Bennett, S., A.~Drummond, D.~Kapan, M.~Suchard, J.~Munoz-Jordan, O.~Pybus,
  E.~Holmes, and D.~Gubler. 2010. Epidemic dynamics revealed in dengue
  evolution. Molecular Biology and Evolution 27:811--818.
\bibAnnoteFile{Bennett2010}

\bibitem[{Bennett et~al.(2003)Bennett, Holmes, Chirivella, Rodriguez, Beltran,
  Vorndam, Gubler, and McMillan}]{Bennett2003}
Bennett, S., E.~Holmes, M.~Chirivella, D.~Rodriguez, M.~Beltran, V.~Vorndam,
  D.~Gubler, and W.~McMillan. 2003. Selection-driven evolution of emergent
  dengue virus. Molecular Biology and Evolution 20:1650--1658.
\bibAnnoteFile{Bennett2003}

\bibitem[{Bhatt et~al.(2013)Bhatt, Gething, Brady, Messina, Farlow, Moyes,
  Drake, Brownstein, Hoen, Sankoh, Myers, George, Jaenisch, Wint, Simmons,
  Scott, Farrar, and Hay}]{Bhatt2013}
Bhatt, S., P.~Gething, O.~Brady, J.~Messina, A.~Farlow, C.~Moyes, J.~Drake,
  J.~Brownstein, A.~Hoen, O.~Sankoh, M.~Myers, D.~George, T.~Jaenisch, G.~Wint,
  C.~Simmons, T.~Scott, J.~Farrar, and S.~Hay. 2013. The global distribution
  and burden of dengue. Nature 496:504--507.
\bibAnnoteFile{Bhatt2013}

\bibitem[{Biek et~al.(2007)Biek, Henderson, Waller, Rupprecht, and
  Real}]{Biek2007}
Biek, R., J.~Henderson, L.~Waller, C.~Rupprecht, and L.~Real. 2007. A
  high-resolution genetic signature of demographic and spatial expansion in
  epizootic rabies virus. Proceedings of the National Academy of Sciences
  104:7993--7998.
\bibAnnoteFile{Biek2007}

\bibitem[{Bloomquist et~al.(2010)Bloomquist, Lemey, and
  Suchard}]{bloomquist2010three}
Bloomquist, E.~W., P.~Lemey, and M.~A. Suchard. 2010. Three roads diverged?
  routes to phylogeographic inference. Trends in Ecology \& Evolution
  25:626--632.
\bibAnnoteFile{bloomquist2010three}

\bibitem[{Brennan et~al.(2008)Brennan, Bodelle, Coffey, Devare, Golden,
  Hackett~Jr., Harris, Holzmayer, Luk, Schochetman, Swanson, Yamaguchi,
  Vallari, Ndembi, Ngansop, Makamche, Mbanya, Gurtler, Zekeng, and
  Kaptue}]{Brennan2008}
Brennan, C., P.~Bodelle, R.~Coffey, S.~Devare, A.~Golden, J.~Hackett~Jr.,
  B.~Harris, V.~Holzmayer, K.~Luk, G.~Schochetman, P.~Swanson, J.~Yamaguchi,
  A.~Vallari, N.~Ndembi, C.~Ngansop, F.~Makamche, D.~Mbanya, L.~Gurtler,
  L.~Zekeng, and L.~Kaptue. 2008. The prevalence of diverse {HIV}-1 strains was
  stable in {C}ameroonian blood donors from 1996 to 2004. Journal of Acquired
  Immune Deficiency Syndrome 49:432--439.
\bibAnnoteFile{Brennan2008}

\bibitem[{Campos et~al.(2010)Campos, Willerslev, Sher, Orlando, Axelsson,
  Tikhonov, Aaris-Sorenson, Greenwood, Kahlke, Kosintsev, Krakhmalnaya,
  Kuznetsova, Lemey, MacPhee, Norris, Shepherd, Suchard, Zazula, Shapiro, and
  Gilbert}]{Campos2010}
Campos, P., E.~Willerslev, A.~Sher, L.~Orlando, E.~Axelsson, A.~Tikhonov,
  K.~Aaris-Sorenson, A.~Greenwood, R.~Kahlke, P.~Kosintsev, T.~Krakhmalnaya,
  T.~Kuznetsova, P.~Lemey, R.~MacPhee, C.~Norris, K.~Shepherd, M.~Suchard,
  G.~Zazula, B.~Shapiro, and M.~Gilbert. 2010. Ancient {DNA} analyses exclude
  humans as the driving force behind late plestocene musk ox (\emph{Ovibos
  moschatus}) population dynamics. Proceedings of the National Academy of
  Sciences 107:5675--5680.
\bibAnnoteFile{Campos2010}

\bibitem[{Carrington et~al.(2005)Carrington, Foster, Pybus, Bennett, and
  Holmes}]{Carrington2005}
Carrington, C., J.~Foster, O.~Pybus, S.~Bennett, and E.~Holmes. 2005. Invasion
  and maintenance of dengue virus type 2 and type 4 in the {A}mericas. Journal
  of Virology 79:14680--14687.
\bibAnnoteFile{Carrington2005}

\bibitem[{Childs et~al.(2000)Childs, Curns, Dey, Real, Feinstein, and
  Bjornstad}]{Childs2000}
Childs, J., A.~Curns, M.~Dey, L.~Real, L.~Feinstein, and O.~Bjornstad. 2000.
  Predicting the local dynamics of epizootic rabies among raccoons in the
  {U}nited {S}tates. Proceedings of the National Academy of Sciences
  97:13666--13671.
\bibAnnoteFile{Childs2000}

\bibitem[{Chipman et~al.(2001)Chipman, George, and McCulloch}]{Chipman2001}
Chipman, H., E.~George, and R.~McCulloch. 2001. The practical implementation of
  {B}ayesian model selection. IMS Lecture Notes - Monograph Series 38:67--134.
\bibAnnoteFile{Chipman2001}

\bibitem[{Crandall et~al.(1999)Crandall, Posada, and Vasco}]{Crandall1999}
Crandall, K., D.~Posada, and D.~Vasco. 1999. Effective population sizes:
  missing measures and missing concepts. Animal Conservation 2:317--319.
\bibAnnoteFile{Crandall1999}

\bibitem[{Cummings et~al.(2004)Cummings, Irizarry, Huang, Endy, Nisalak,
  Ungchusak, and Burke}]{Cummings2004}
Cummings, D., R.~Irizarry, N.~Huang, T.~Endy, A.~Nisalak, K.~Ungchusak, and
  D.~Burke. 2004. Travelling waves in the occurrence of dengue haemorrhagic
  fever in {T}hailand. Nature 427:344--347.
\bibAnnoteFile{Cummings2004}

\bibitem[{Dansgaard et~al.(1993)Dansgaard, Johnsen, Clausen, Dahl-Jensen,
  Gundestrup, Hammer, Hvidberg, Steffensen, Sveinbjornsdottir, Jouzel, and
  Bond}]{Dansgaard1993}
Dansgaard, W., S.~Johnsen, H.~Clausen, D.~Dahl-Jensen, N.~Gundestrup,
  C.~Hammer, C.~Hvidberg, J.~Steffensen, A.~Sveinbjornsdottir, J.~Jouzel, and
  G.~Bond. 1993. Evidence for general instability of past climate from a 250
  kyr ice-core record. Nature 364:218--220.
\bibAnnoteFile{Dansgaard1993}

\bibitem[{Dansgaard et~al.(1989)Dansgaard, White, and Johnsen}]{Dansgaard1989}
Dansgaard, W., J.~White, and S.~Johnsen. 1989. The abrupt termination of the
  {Y}ounger {D}ryas climate event. Nature 339:532--533.
\bibAnnoteFile{Dansgaard1989}

\bibitem[{Donnelly and Tavar\'{e}(1995)}]{Donnelly1995}
Donnelly, P. and S.~Tavar\'{e}. 1995. Coalescents and genealogical structure
  under neutrality. Annual Review of Genetics 29:401--421.
\bibAnnoteFile{Donnelly1995}

\bibitem[{Drummond et~al.(2002)Drummond, Nicholls, Rodrigo, and
  Solomon}]{Drummond2002}
Drummond, A., G.~Nicholls, A.~Rodrigo, and W.~Solomon. 2002. Estimating
  mutation parameters, population history and genealogy simultaneously from
  temporally spaced sequence data. Genetics 161:1307--1320.
\bibAnnoteFile{Drummond2002}

\bibitem[{Drummond et~al.(2005)Drummond, Rambaut, Shapiro, and
  Pybus}]{Drummond2005}
Drummond, A., A.~Rambaut, B.~Shapiro, and O.~Pybus. 2005. Bayesian coalescent
  inference of past population dynamics from molecular sequences. Molecular
  Biology and Evolution 22:1185--1192.
\bibAnnoteFile{Drummond2005}

\bibitem[{Drummond et~al.(2012)Drummond, Suchard, Xie, and
  Rambaut}]{Drummond2012}
Drummond, A.~J., M.~A. Suchard, D.~Xie, and A.~Rambaut. 2012. Bayesian
  phylogenetics with beauti and the beast 1.7. Molecular biology and evolution
  29:1969--1973.
\bibAnnoteFile{Drummond2012}

\bibitem[{Edo-Matas et~al.(2011)Edo-Matas, Lemey, Tom, Serna-Bolea, van~den
  Blink, van't Wout, Schuitemaker, and Suchard}]{Edo-Matas2011}
Edo-Matas, D., P.~Lemey, J.~A. Tom, C.~Serna-Bolea, A.~E. van~den Blink, A.~B.
  van't Wout, H.~Schuitemaker, and M.~A. Suchard. 2011. Impact of {CCR5delta32}
  host genetic background and disease progression on {HIV}-1 intrahost
  evolutionary processes: efficient hypothesis testing through hierarchical
  phylogenetic models. Molecular Biology and Evolution 28:1605--16.
\bibAnnoteFile{Edo-Matas2011}

\bibitem[{Faria et~al.(2014)Faria, Rambaut, Suchard, Baele, Bedford, Ward,
  Tatem, Sousa, Arinaminpathy, Pepin, Posada, Peeters, Pybus, and
  Lemey}]{Faria2014}
Faria, N., A.~Rambaut, M.~Suchard, G.~Baele, T.~Bedford, M.~Ward, A.~Tatem,
  J.~Sousa, N.~Arinaminpathy, J.~Pepin, D.~Posada, M.~Peeters, O.~Pybus, and
  P.~Lemey. 2014. The early spread and epidemic ignition of {HIV-1} in human
  populations. Science 346:56--61.
\bibAnnoteFile{Faria2014}

\bibitem[{Faria et~al.(2012)Faria, Suchard, Abecasis, Sousa, Ndembi, Bonfim,
  Camacho, Vandamme, and Lemey}]{Faria2011}
Faria, N., M.~Suchard, A.~Abecasis, J.~Sousa, N.~Ndembi, I.~Bonfim, R.~Camacho,
  A.~Vandamme, and P.~Lemey. 2012. Phylodynamics of the {HIV}-1 {CRF}02\_{AG}
  clade in {C}ameroon. Infection, Genetics and Evolution 12:453--460.
\bibAnnoteFile{Faria2011}

\bibitem[{Finlay et~al.(2007)Finlay, Gaillard, Vahidi, Mirhoseini, Jianlin, Qi,
  El-Barody, Baird, Healy, and Bradley}]{Finlay2007}
Finlay, E., C.~Gaillard, S.~Vahidi, S.~Mirhoseini, H.~Jianlin, X.~Qi,
  M.~El-Barody, J.~Baird, B.~Healy, and D.~Bradley. 2007. Bayesian inference of
  population expansions in domestic bovines. Biology Letters 3:449--452.
\bibAnnoteFile{Finlay2007}

\bibitem[{Frost and Volz(2010)}]{Frost2010}
Frost, S. and E.~Volz. 2010. Viral phylodynamics and the search for an
  {`}effective number of infections{'}. Philosophical Transactions of the Royal
  Society B 365:1879--1890.
\bibAnnoteFile{Frost2010}

\bibitem[{George and McCulloch(1993)}]{George1993}
George, E. and R.~McCulloch. 1993. Variable selection via {G}ibbs sampling.
  Journal of American Statistical Association 88:881--889.
\bibAnnoteFile{George1993}

\bibitem[{Gill et~al.(2013)Gill, Lemey, Faria, Rambaut, Shapiro, and
  Suchard}]{Gill2013}
Gill, M., P.~Lemey, N.~Faria, A.~Rambaut, B.~Shapiro, and M.~Suchard. 2013.
  Improving {B}ayesian population dynamics inference: a coalescent-based model
  for multiple loci. Molecular Biology and Evolution 30:713--724.
\bibAnnoteFile{Gill2013}

\bibitem[{Griffiths and Tavar\'{e}(1994)}]{Griffiths1994}
Griffiths, R. and S.~Tavar\'{e}. 1994. Sampling theory for neutral alleles in a
  varying environment. Philosophical Transactions of the Royal Society of
  London. Series B, Biological Sciences 344:403--410.
\bibAnnoteFile{Griffiths1994}

\bibitem[{{{{GRIP} {M}embers}}(1993)}]{GRIP1993}
{{{GRIP} {M}embers}}. 1993. Climate instability during the last interglacial
  period recorded in the {GRIP} ice core. Nature 364:203--207.
\bibAnnoteFile{GRIP1993}

\bibitem[{Grootes et~al.(1993)Grootes, Stuiver, White, Johnsen, and
  Jouzel}]{Grootes1993}
Grootes, P., M.~Stuiver, J.~White, S.~Johnsen, and J.~Jouzel. 1993. Comparison
  of oxygen isotope records from the {GISP2} and {GRIP} {G}reenland ice cores.
  Nature 366:552--554.
\bibAnnoteFile{Grootes1993}

\bibitem[{Hastings(1970)}]{Hastings1970}
Hastings, W. 1970. {M}onte {C}arlo sampling methods using {M}arkov chains and
  their applications. Biometrika 57:97--109.
\bibAnnoteFile{Hastings1970}

\bibitem[{Heled and Drummond(2008)}]{Heled2008}
Heled, J. and A.~Drummond. 2008. Bayesian inference of population size history
  from multiple loci. BMC Evolutionary Biology 8:289.
\bibAnnoteFile{Heled2008}

\bibitem[{Hemelaar et~al.(2011)Hemelaar, Gouws, Ghys, Osmanov, and {WHO-UNAIDS
  Network for HIV Isolation and Characterisation}}]{Hemelaar:2011fk}
Hemelaar, J., E.~Gouws, P.~D. Ghys, S.~Osmanov, and {WHO-UNAIDS Network for HIV
  Isolation and Characterisation}. 2011. Global trends in molecular
  epidemiology of {HIV}-1 during 2000-2007. AIDS 25:679--89.
\bibAnnoteFile{Hemelaar:2011fk}

\bibitem[{Ho and Shapiro(2011)}]{Ho2011}
Ho, S. and B.~Shapiro. 2011. Skyline-plot methods of estimating demographic
  history from nucleotide sequences. Molecular Ecology Resources 11:423--434.
\bibAnnoteFile{Ho2011}

\bibitem[{Hudson(1983)}]{Hudson1983}
Hudson, R. 1983. Properties of a neutral allele model with intragenic
  recombination. Theoretical Population Biology 23:183--201.
\bibAnnoteFile{Hudson1983}

\bibitem[{Johnsen et~al.(1997)Johnsen, Clausen, Dansgaard, Gundestrup, Hammer,
  Andersen, Andersen, Hvidberg, Dahl-Jensen, Steffensen, Shoji,
  Sveinbjornsdottir, White, Jouzel, and Fisher}]{Johnsen1997}
Johnsen, S., H.~Clausen, W.~Dansgaard, N.~Gundestrup, C.~Hammer, U.~Andersen,
  K.~Andersen, C.~Hvidberg, D.~Dahl-Jensen, J.~Steffensen, H.~Shoji,
  A.~Sveinbjornsdottir, J.~White, J.~Jouzel, and D.~Fisher. 1997. The d18{O}
  record along the {G}reenland {I}ce {C}ore {P}roject deep ice core and the
  problem of possible {E}emian climatic instability. Journal of Geophysical
  Research 102:26397--26410.
\bibAnnoteFile{Johnsen1997}

\bibitem[{Kappus et~al.(1970)Kappus, Bigler, McLean, and Trevino}]{Kappus1970}
Kappus, K., W.~Bigler, R.~McLean, and H.~Trevino. 1970. The raccoon as an
  emerging rabies host. Journal of Wildlife Diseases 6:507--509.
\bibAnnoteFile{Kappus1970}

\bibitem[{Kingman(1982{\natexlab{a}})}]{Kingman1982b}
Kingman, J. 1982{\natexlab{a}}. The coalescent. Stochastic Processes and their
  Applications 13:235--248.
\bibAnnoteFile{Kingman1982b}

\bibitem[{Kingman(1982{\natexlab{b}})}]{Kingman1982}
Kingman, J. 1982{\natexlab{b}}. On the genealogy of large populations. Journal
  of Applied Probability 19:27--43.
\bibAnnoteFile{Kingman1982}

\bibitem[{Knorr-Held and Rue(2002)}]{Knorr-Held2002}
Knorr-Held, L. and H.~Rue. 2002. On block updating in {M}arkov random field
  models for desease mapping. Scandinavian Journal of Statistics 29:597--614.
\bibAnnoteFile{Knorr-Held2002}

\bibitem[{Krone and Neuhauser(1997)}]{Krone1997}
Krone, S. and C.~Neuhauser. 1997. Ancestral processes with selection.
  Theoretical Population Biology 51:210--237.
\bibAnnoteFile{Krone1997}

\bibitem[{Kuhner et~al.(1998)Kuhner, Yamato, and Felsenstein}]{Kuhner1998}
Kuhner, M., J.~Yamato, and J.~Felsenstein. 1998. Maximum likelihood estimation
  of population growth rates based on the coalescent. Genetics 149:429--434.
\bibAnnoteFile{Kuhner1998}

\bibitem[{Kuo and Mallick(1998)}]{Kuo1998}
Kuo, L. and B.~Mallick. 1998. Variable selection for regression models. Sankhya
  B 60:65--81.
\bibAnnoteFile{Kuo1998}

\bibitem[{Lemey et~al.(2003)Lemey, Pybus, Wang, Saksena, Salemi, and
  Vandamme}]{Lemey2003}
Lemey, P., O.~Pybus, B.~Wang, N.~Saksena, M.~Salemi, and A.~Vandamme. 2003.
  Tracing the origin and history of the {HIV}-2 epidemic. Proceedings of the
  National Academy of Sciences 100:6588--6592.
\bibAnnoteFile{Lemey2003}

\bibitem[{Liu and Mittler(2008)}]{Liu2008}
Liu, Y. and J.~Mittler. 2008. Selection dramatically reduces effective
  population size in {HIV}-1 infection. BMC Evolutionary Biology 8:133.
\bibAnnoteFile{Liu2008}

\bibitem[{Lorenzen et~al.(2011)Lorenzen, Nogues-Braco, Orlando, Weinstock,
  Binladen, Marske, Ugan, Borregaard, Gilbert, Nielsen, Ho, Goebel, Graf,
  Byers, Stenderup, Rasmussen, Campos, Leonard, Koepfli, Froese, Zazula,
  Stafford, Aaris-Sorensen, Batra, Haywood, Singarayer, Valdes, Boeskorov,
  Burns, Davydov, Haile, Jenkins, Kosintsev, Kuznetsova, Lai, Martin, McDonald,
  Mol, Meldgaard, Munch, Stephan, Sablin, Sommer, Sipko, Scott, Suchard,
  Tikhonov, Willerslev, Wayne, Cooper, Hofreiter, Sher, Shapiro, Rahbek, and
  Willerslev}]{Lorenzen2011}
Lorenzen, E., D.~Nogues-Braco, L.~Orlando, J.~Weinstock, J.~Binladen,
  K.~Marske, A.~Ugan, M.~Borregaard, M.~Gilbert, R.~Nielsen, S.~Ho, T.~Goebel,
  K.~Graf, D.~Byers, J.~Stenderup, M.~Rasmussen, P.~Campos, J.~Leonard,
  K.~Koepfli, D.~Froese, G.~Zazula, T.~Stafford, K.~Aaris-Sorensen, P.~Batra,
  A.~Haywood, J.~Singarayer, P.~Valdes, G.~Boeskorov, J.~Burns, S.~Davydov,
  J.~Haile, D.~Jenkins, P.~Kosintsev, T.~Kuznetsova, X.~Lai, L.~Martin,
  H.~McDonald, D.~Mol, M.~Meldgaard, K.~Munch, E.~Stephan, M.~Sablin,
  R.~Sommer, T.~Sipko, E.~Scott, M.~Suchard, A.~Tikhonov, R.~Willerslev,
  R.~Wayne, A.~Cooper, M.~Hofreiter, A.~Sher, B.~Shapiro, C.~Rahbek, and
  E.~Willerslev. 2011. Species-specific responses of {L}ate {Q}uaternary
  megafauna to climate and humans. Nature 479:359--365.
\bibAnnoteFile{Lorenzen2011}

\bibitem[{Metropolis et~al.(1953)Metropolis, Rosenbluth, Rosenbluth, Teller,
  and Teller}]{Metropolis1953}
Metropolis, N., A.~Rosenbluth, M.~Rosenbluth, A.~Teller, and E.~Teller. 1953.
  Equation of state calculation by fast computing machines. Journal of Chemical
  Physics 21:1087--1092.
\bibAnnoteFile{Metropolis1953}

\bibitem[{Minin et~al.(2008)Minin, Bloomquist, and Suchard}]{Minin2008}
Minin, V., E.~Bloomquist, and M.~Suchard. 2008. Smooth skyride through a rough
  skyline: Bayesian coalescent based inference of population dynamics.
  Molecular Biology and Evolution 25:1459--1471.
\bibAnnoteFile{Minin2008}

\bibitem[{Notohara(1990)}]{Notohara1990}
Notohara, M. 1990. The coalescent and the genealogical process in
  geographically structured population. Journal of Mathematical Biology
  29:59--75.
\bibAnnoteFile{Notohara1990}

\bibitem[{Opgen-Rhein et~al.(2005)Opgen-Rhein, Fahrmeir, and
  Strimmer}]{Opgen-Rhein2005}
Opgen-Rhein, R., L.~Fahrmeir, and K.~Strimmer. 2005. Inference of demographic
  history from genealogical trees using reversible jump {M}arkov chain {M}onte
  {C}arlo. BMC Evolutionary Biology 5:6.
\bibAnnoteFile{Opgen-Rhein2005}

\bibitem[{Palacios and Minin(2013)}]{Palacios2013}
Palacios, J. and V.~Minin. 2013. Gaussian process-based {B}ayesian
  nonparametric inference of population size trajectories from gene
  genealogies. Biometrics 69:8--18.
\bibAnnoteFile{Palacios2013}

\bibitem[{Palstra and Fraser(2012)}]{Palstra2012}
Palstra, F. and D.~Fraser. 2012. Effective/census population size ratio
  estimation: a compendium and appraisal. Ecology and Evolution 2:2357--2365.
\bibAnnoteFile{Palstra2012}

\bibitem[{Powell et~al.(2010)Powell, Barengolts, Mayr, and Nyambi}]{Powell2010}
Powell, R., D.~Barengolts, L.~Mayr, and P.~Nyambi. 2010. The evolution of
  {HIV}-1 diversity in rural {C}ameroon and its implications in vaccine design
  and trials. Viruses 2:639--654.
\bibAnnoteFile{Powell2010}

\bibitem[{Pybus et~al.(2000)Pybus, Rambaut, and Harvey}]{Pybus2000}
Pybus, O., A.~Rambaut, and P.~Harvey. 2000. An integrated framework for the
  inference of viral population history from reconstructed genealogies.
  Genetics 155:1429--1437.
\bibAnnoteFile{Pybus2000}

\bibitem[{Rodrigo and Felsenstein(1999)}]{Rodrigo1999b}
Rodrigo, A. and J.~Felsenstein. 1999. Coalescent Approaches to {HIV} Population
  Genetics Pages~233--274. Johns Hopkins Universtiy Press, Baltimore, MD.
\bibAnnoteFile{Rodrigo1999b}

\bibitem[{Shapiro et~al.(2004)Shapiro, Drummond, Rambaut, Wilson, Matheus,
  Sher, Pybus, Gilbert, Barnes, Binladen, Willerslev, Hansen, Baryshnikov,
  Burns, Davydov, Driver, Froese, Harington, Keddie, Kosintsev, Kunz, Martin,
  Stephenson, Storer, Tedford, Zimov, and Cooper}]{Shapiro2004}
Shapiro, B., A.~Drummond, A.~Rambaut, M.~Wilson, P.~Matheus, A.~Sher, O.~Pybus,
  M.~Gilbert, I.~Barnes, J.~Binladen, E.~Willerslev, A.~Hansen, G.~Baryshnikov,
  J.~Burns, S.~Davydov, J.~Driver, D.~Froese, C.~Harington, G.~Keddie,
  P.~Kosintsev, M.~Kunz, L.~Martin, R.~Stephenson, J.~Storer, R.~Tedford,
  S.~Zimov, and A.~Cooper. 2004. Rise and fall of the {B}eringian steppe bison.
  Science 306:1561--1565.
\bibAnnoteFile{Shapiro2004}

\bibitem[{Slatkin and Hudson(1991)}]{Slatkin1991}
Slatkin, M. and R.~Hudson. 1991. Pairwise comparison of mitochondrial {DNA}
  sequences in stable and exponentially growing populations. Genetics
  129:555--562.
\bibAnnoteFile{Slatkin1991}

\bibitem[{Stiller et~al.(2010)Stiller, Baryshnikov, Bocherens,
  Grandal-d'Anglade, Hilpert, Munzel, Pinhasi, Rabeder, Rosendahl, Trinkaus,
  Hofreiter, and Knapp}]{Stiller2010}
Stiller, M., G.~Baryshnikov, H.~Bocherens, A.~Grandal-d'Anglade, B.~Hilpert,
  S.~Munzel, R.~Pinhasi, G.~Rabeder, W.~Rosendahl, E.~Trinkaus, M.~Hofreiter,
  and M.~Knapp. 2010. Withering away-25,000 years of genetic decline preceded
  cave bear extinction. Molecular Biology and Evolution 27:975--978.
\bibAnnoteFile{Stiller2010}

\bibitem[{Strimmer and Pybus(2001)}]{Strimmer2001}
Strimmer, K. and O.~Pybus. 2001. Exploring the demographic history of {DNA}
  sequences using the generalized skyline plot. Molecular Biology and Evolution
  18:2298--2305.
\bibAnnoteFile{Strimmer2001}

\bibitem[{Stuart et~al.(2004)Stuart, Kosintsev, Higham, and
  Lister}]{Stuart2004}
Stuart, A., P.~Kosintsev, T.~Higham, and A.~Lister. 2004. Pleistocene to
  {H}olocene extinction dynamics in giant deer and wooly mammoth. Nature
  431:684--689.
\bibAnnoteFile{Stuart2004}

\bibitem[{Suchard et~al.(2003)Suchard, Kitchen, Sinsheimer, and
  Weiss}]{Suchard2003b}
Suchard, M., C.~Kitchen, J.~Sinsheimer, and R.~Weiss. 2003. Hierarchical
  phylogenetic models for analyzing multipartite sequence data. Systematic
  Biology 52:649--664.
\bibAnnoteFile{Suchard2003b}

\bibitem[{Tener(1965)}]{Tener1965}
Tener, J. 1965. Muskoxen in {C}anada: a biological and taxonomic review.
  Wildlife Service Monograph Series No. 2 .
\bibAnnoteFile{Tener1965}

\bibitem[{UNAIDS(2015)}]{UNAIDS}
UNAIDS. 2015. {AIDS}info. \url{http://aidsinfo.unaids.org/}.
\bibAnnoteFile{UNAIDS}

\bibitem[{Volz et~al.(2009)Volz, Pond, Ward, Brown, and Frost}]{Volz2009}
Volz, E., S.~K. Pond, M.~Ward, A.~L. Brown, and S.~Frost. 2009. Phylodynamics
  of infectious disease epidemics. Genetics 183:1421--1430.
\bibAnnoteFile{Volz2009}

\bibitem[{WHO(2015{\natexlab{a}})}]{WHODengue}
WHO. 2015{\natexlab{a}}. World {H}ealth {O}rganization, {D}engue.
  \url{http://www.who.int/topics/dengue/en/}.
\bibAnnoteFile{WHODengue}

\bibitem[{WHO(2015{\natexlab{b}})}]{WHORabies}
WHO. 2015{\natexlab{b}}. World {H}ealth {O}rganization, {R}abies.
  \url{http://www.who.int/rabies/en/}.
\bibAnnoteFile{WHORabies}

\bibitem[{Wright(1931)}]{Wright1931}
Wright, S. 1931. Evolution in {M}endelian populations. Genetics 16:97--159.
\bibAnnoteFile{Wright1931}

\end{thebibliography}

\end{document}